\documentclass[11pt,letterpaper]{article}

\usepackage[pdftex]{graphicx}
\usepackage{amssymb}
\usepackage{amsmath}
\usepackage{amsfonts}
\usepackage{appendix}
\usepackage{multirow}
\usepackage{color}
\usepackage{url}
\usepackage{caption}
\usepackage{pdflscape}
\usepackage{afterpage}
\usepackage{graphicx}
\usepackage{longtable}
\usepackage{booktabs, caption, makecell}
\usepackage{threeparttable}
\usepackage{pdfpages}
\usepackage{chngpage}
\usepackage[normalem]{ulem}
\usepackage{cancel}
\usepackage{bm}
\usepackage{rotating}

\newtheorem{defin}{Definition}

\newtheorem{prop}{Proposition}
\newtheorem{rem}{Remark}
\newtheorem{theo}{Theorem}

\begin{document}

\title{\textsc{Discovery and Equilibrium in Games with Unawareness\thanks{I thank the editor, Marciano Siniscalchi, and two anonymous reviewers for their constructive comments. Moreover, I thank Aviad Heifetz, Byung Soo Lee, and seminar participants at UC Berkeley, the University of Toronto, the Barcelona workshop on Limited Reasoning and Cognition, SAET 2015, CSLI 2016, TARK 2017, and the Virginia Tech Workshop on Advances in Decision Theory 2018 for helpful discussions. An abbreviated earlier version appeared in the online proceedings of TARK 2017 under the title ``Self-Confirming Games: Unawareness, Discovery, and Equilibrium''. I am grateful for past financial support through NSF SES-0647811.}}}

\author{Burkhard C. Schipper\thanks{Department of Economics, University of California, Davis. Email: bcschipper@ucdavis.edu}}

\date{September 10, 2021}

\maketitle

\begin{abstract} Equilibrium notions for games with unawareness in the literature cannot be interpreted as steady-states of a learning process because players may discover novel actions during play. In this sense, many games with unawareness are ``self-destroying'' as a player's representation of the game may change after playing it once. We define discovery processes where at each state there is an extensive-form game with unawareness that together with the players' play determines the transition to possibly another extensive-form game with unawareness in which players are now aware of actions that they have discovered. A discovery process is rationalizable if players play extensive-form rationalizable strategies in each game with unawareness. We show that for any game with unawareness there is a rationalizable discovery process that leads to a self-confirming game that possesses a self-confirming equilibrium in extensive-form rationalizable conjectures. This notion of equilibrium can be interpreted as steady-state of both a discovery and learning process.
\bigskip


\bigskip

\noindent \textbf{JEL-Classifications: } C72, D83.
\end{abstract}

\thispagestyle{empty}

\pagenumbering{empty}

\renewcommand{\baselinestretch}{1.2}

\small\normalsize

\pagenumbering{arabic}

\newpage

\section{Introduction}

How do players arrive at their conceptions of a strategic situation? Game theory is mostly concerned with finding optimal behavior \emph{given} a formal representation of the strategic situation. However, where do player's representations of the strategic situation come from? Have they been discovered during earlier strategic interaction? If this is the case, then the players' views of the strategic situation should be the result of strategic interaction rather than an assumption. This is the main issue attacked in this paper. This view leads to further questions such as `Need representations of the strategic situation be necessarily common among all players as it is assumed in standard game theory?' Even in standard games of incomplete information the description of the strategic situation including all relevant uncertainties is shared among all players (and the analyst) and is thus common to all players. Players may have different information but all players conceive of the same set of uncertainties, actions etc.

Previously, game theory has been criticized as a formal apparatus that is incapable of modeling novelty, discovery, and surprise. For instance, Shackle (1972, p. 161) wrote ``The Theory of Games thus supposes the players to have knowledge of all the possibilities: \emph{surprise}, the most powerful and incise element in the whole art of war, is eliminated by the theoretical frame itself; and novelty, the \emph{changing} of what appeared to be the roles of the game, the continually threatening dissolution of the conditions and circumstances in which either player may suppose himself to be operating, is eliminated also, by the supposition that each player, like a chess player of super-human intellectual range, knows everything that can happen.'' We aim to demonstrate that game theory with unawareness is sufficiently rich for modelling novelty, surprise, transformative experiences (Paul, 2014), discoveries, shattering of player's views of the strategic situation etc.

This paper is inspired by the literature on unawareness in games. In particular, our motivation is the quest for a natural notion of equilibrium to games with unawareness. Various frameworks for modeling dynamic games with unawareness have been introduced (Halpern and Rego, 2014, Rego and Halpern, 2012, Feinberg, 2020, Li, 2008, Grant and Quiggin, 2013, Heifetz, Meier, and Schipper, 2013; for a non-technical survey, see Schipper, 2014). While all of those frameworks are capable of modeling strategic interaction under asymmetric unawareness at various degrees of generality and tractability, the solution concepts proposed for those frameworks and thus the implicit behavioral assumptions under unawareness differ. The solution concepts that have been proposed in the literature can roughly be divided into equilibrium notions (Halpern and Rego, 2014, Rego and Halpern, 2012, Feinberg, 2020, Li 2008, Grant and Quiggin, 2013, Ozbay, 2007, Meier and Schipper, 2013) and rationalizability notions (Heifetz, Meier, and Schipper, 2013, 2021, Meier and Schipper, 2012, Perea, 2021). Authors proposing equilibrium notions to dynamic games with unawareness appear to be mainly guided by extending the mathematical definitions of equilibrium in standard games to the more sophisticated frameworks with unawareness. Yet, I believe less attention has been paid to the interpretations of the behavioral assumptions embodied in these standard equilibrium concepts and whether or not such interpretations could meaningfully extent also to dynamic games with unawareness. Applied work featuring asymmetric unawareness used both the ``equilibrium'' approach (e.g, Filiz-Ozbay, 2012, van Thadden and Zhao, 2012, Auster, 2013, Auster and Pavoni, 2021, Schipper and Zhou, 2021) and the rationalizability approach (e.g., Schipper and Woo, 2019, Francetich and Schipper, 2021). But the interpretation of the results in these applications especially with respect to the implicit time horizon of the prediction depends crucially on the interpretation of the solution concept.

In standard game theory, equilibrium is interpreted as an outcome in which each player plays ``optimally'' given the opponents' play. It features not just the rationality assumption but also mutual knowledge of play. This mutual knowledge of play may emerge in a steady-state of a learning process (Fudenberg and Levine, 1998, Foster and Young, 2003, Hart and Mas-Colell, 2006). This interpretation cannot apply generally to games with unawareness. This is because players may be unaware of actions and may discover novel actions during play. The ''next time'' they play ``the game'', they actually play a different game in which now they are aware of previously discovered actions. That is, dynamic learning processes in games with unawareness must not only deal with learning about opponents' play but also with \emph{discoveries} that may lead to transformative changes in players' views of the game.\footnote{The conceptual difference between learning and discovery is roughly as follows: When a player learns, she discards possibilities. When a player discovers, she adds possibilities that she had not previously conceived.} Games with unawareness may be ``self-destroying'' representations of the strategic situation in the sense that rational play may destroy some player's view of the strategic situation. Only when a view of the strategic situation is ``self-confirming'', i.e., rational play in such a game does not lead to further changes in the players' views of the game, an equilibrium notion as a steady-state of a learning process of behavior may be meaningfully applied. Our paper seeks to make this precise.

We introduce a notion of self-confirming equilibrium for extensive-form games with unawareness. In self-confirming equilibrium, nobody discovers that their own view of the game may be incomplete. Moreover, players play optimally in this extensive-form game given their beliefs and their beliefs are not falsified by their play. Self-confirming equilibrium may fail to exist in an extensive-form game with unawareness because rational play may lead to discoveries. We formalize the notion of discovered game: For any extensive-form game with unawareness and strategy profile, the discovered game is a game in which each player's awareness is ``updated'' given their discoveries but their information stays essentially the same (modulo awareness). This leads to a notion akin to stochastic games except that states correspond now to extensive-form games with unawareness and the transition probabilities model for each extensive-form game with unawareness and strategy profile the transition to the discovered game. Such a stochastic game and a Markov strategy that assigns to each extensive-form game with unawareness a mode of behavior we call a discovery process. We assume that at each state players just optimize within the extensive-form game associated with the state instead of over all extensive-form games across the entire discovery process.

We select among discovery processes by requiring the Markov strategy in the stochastic game to assign only extensive-form rationalizable strategies to each extensive-form game with unawareness. For every finite extensive-form game with unawareness, there exists an extensive-form rationalizable discovery process that leads to an extensive-form game with unawareness that is an absorbing state of the process. We consider it as a steady-state of conceptions when players play at each state of the stochastic game with common (strong) belief in rationality and call it the rationalizable self-confirming game. In such a game, it makes sense to look also for a steady-state of a learning process of behavior. The natural outcome of such a learning process is a self-confirming equilibrium (Battigalli, 1987, Fudenberg and Levine, 1993a, Kalai and Lehrer, 1993b). Moreover, since we assumed that players play extensive-form rationalizable strategies all along the discovery process, it makes also sense to focus on self-confirming equilibrium involving only extensive-form rationalizable strategies, a notion of equilibrium that has been previously discussed in an example of a macroeconomic game by Battigalli and Guaitoli (1997) and has recently been studied in more detail by Battigalli and Bordoli (2020). Essentially we show an existence result for equilibrium in games with unawareness: We observe that for every extensive-form game with unawareness there exists a rationalizable discovery process leading to a rationalizable self-confirming game that possesses a self-confirming equilibrium in extensive-form rationalizable conjectures. This is a notion of equilibrium both in terms of conceptions of the strategic situation as well as strategic behavior.

Before we proceed with our formal exposition, we should clarify some methodological aspects upfront: First, we focus on the discovery process rather than the learning process. Although we motivate our solution concept by a learning and discovery process of recurrent play of the strategic situation, we do not formally model the learning process allowing behavior to converge once a self-confirming game has been reached in the discovery process. Such learning processes have been studied (for games without unawareness) by Fudenberg and Levine (1993b) for their notion of self-confirming equilibrium and recently by Battigalli and Bordoli (2020) for self-confirming equilibrium in extensive-form rationalizable conjectures. We instead focus here on the discovery process because we believe that this is the novel contribution to the theory of games and the literature on unawareness. Second, we consider a fixed set of players rather than a population of players with random matching after each stage-game. If we were to allow random matching among a population of players, each player would also need to keep track of the distribution of awareness in the population (bounded by her own awareness). In order to avoid this complication but nevertheless rule out repeated games effects as well as strategic learning and teaching across games along the discovery process, we assume that players are impatient and just maximize expected payoffs \emph{within} each stage-game rather than maximizing expected payoffs intertemporally across games along the discovery process. This is because our main goal is to seek a justification for a notion of ``stage-game'' equilibrium under unawareness rather than any other equilibria of repeated games. This approach is also taken by Battigalli and Bordoli (2020) for their learning justification of self-confirming stage-game equilibrium in extensive-form rationalizable conjectures in games without unawareness.\footnote{The assumption of impatient players without random matching became somewhat common in the recent literature on learning stage-game equilibrium (e.g. Hart and Mas-Colell, 2006). Learning stage-game equilibrium can be in jeopardy if strategic teaching by a long-run player is allowed (see Schipper, 2020).} Third, we consider extensive-form games with unawareness rather than just normal-form games with unawareness because former allow for a richer consideration of what could be rationally discovered in a game. We view the definition of updated information sets of discovered versions as one of the main contributions of the paper. Moreover, extensive-form games allow us to consider a strong refinement of self-confirming equilibrium by extensive-form rationalizable conjectures that are known to involve forward induction. Note that games with unawareness in normal-form are a special case of our analysis as we allow for simultaneous moves. Finally, we should mention that while our analysis focuses on unawareness proper, we outline in Section~\ref{awunaw} how the framework can be extended to also encompass awareness of unawareness and the additional issues that arise under awareness of unawareness.

The paper is organized as follow: We illustrate our approach with simple examples in the next section. In Section~\ref{model} we introduce the basic model of extensive-form games and unawareness and self-confirming equilibrium. This is followed in Section~\ref{discovery_section} with (extensive-form) discovery processes and (rationalizable) self-confirming games admitting a self-confirming equilibrium (in extensive-form rationalizable conjectures). Finally, Section~\ref{discussion} concludes with a discussion including of the related literature. Proofs are relegated to the appendix.

\section{Simple Illustrating Examples\label{Example}}

\noindent \textbf{Example 1 } The first simple example illustrates that any definition of equilibrium in the literature on unawareness misses the essence of what is equilibrium. It cannot be interpreted as a steady-state of behavior. The example also illustrates some features of our framework.
\begin{figure}[h!]
\caption{(Initial) Game of Example 1}
\label{example1a}
\begin{center}
\includegraphics[scale=0.35]{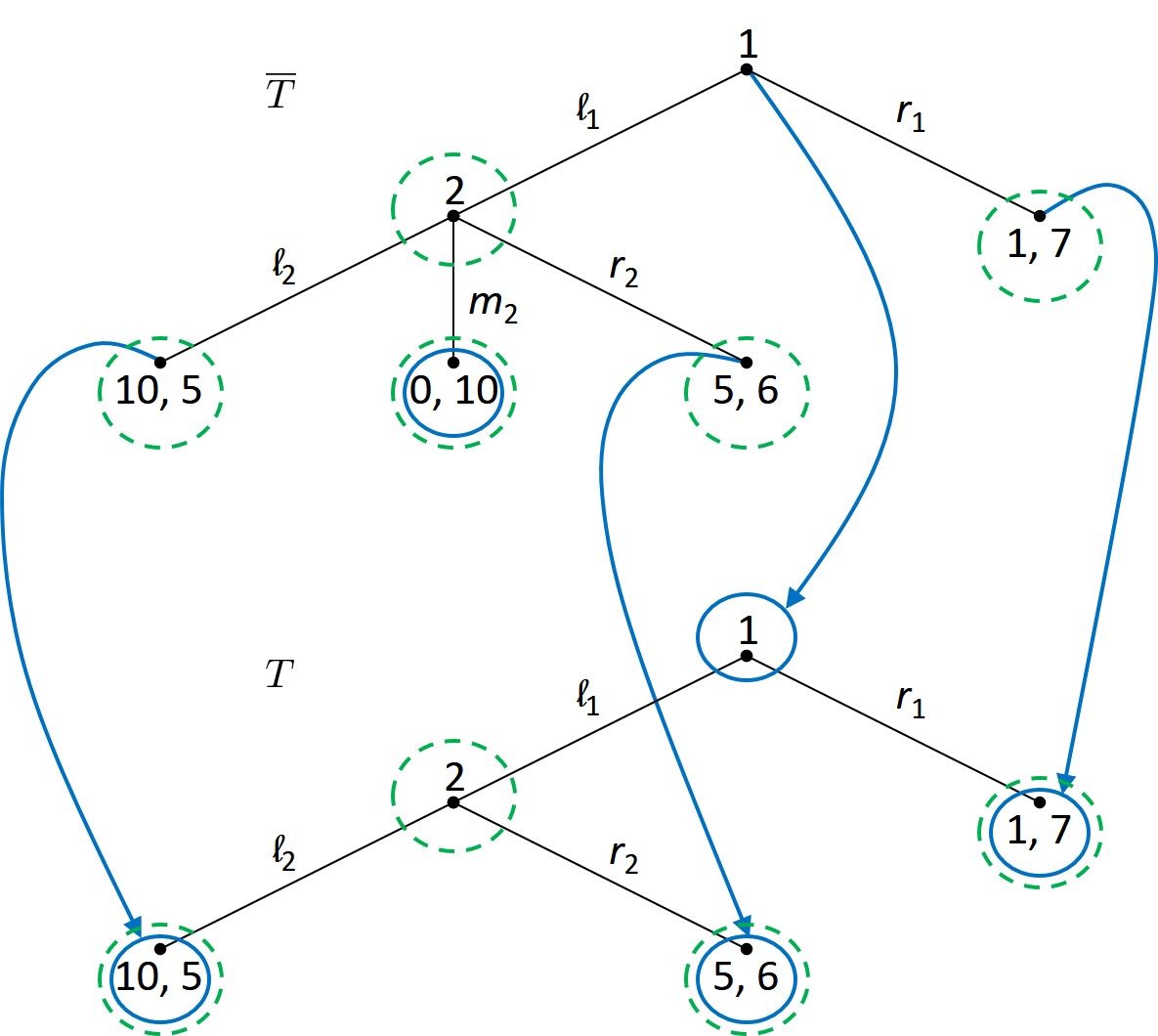}
\end{center}
\end{figure}

There are two players, 1 and 2. Player 1 (e.g., the principal) moves first. She can either delegate to player 2 (e.g., an agent) or do the work by herself. In the latter case, the game ends and both players receive their payoffs. If player 1 delegates to player 2, then player 2 can take one out of three actions. So far, it sounds like a most basic two-stage principal-agent problem. The non-standard but straightforward detail is that player 1 is not aware of all of player 2's actions (and does not even realize this). She considers only two actions of player 2. This strategic situation is modeled in the game depicted in Figure~\ref{example1a}.

There are two trees. The tree at the bottom, $T$, is a subtree of the tree at the top, $\bar{T}$, in the sense that action $m_2$ of player 2 is missing in $T$. This illustrates one non-standard feature of games with unawareness, namely that instead of just one tree we consider a forest of trees that differ in how ``rich'' they describe the situation. The information and awareness of both players are modeled with information sets. The solid-lined blue spheres and arrows belong to player 1, the dashed green spheres belong to player 2. At any node in which a player is active, the player is aware of the tree on which her information set at this node is located. There are two non-standard features of these information sets: First, the information set of a decision node in one tree may consist of decision nodes in a lower tree $T$. For instance, player 1's information set at the beginning of the game in the upper tree $\bar{T}$ is in the lower tree $T$. This signifies the fact that initially player 1 is unaware of player 2's action $m_2$ and thus considers the strategic situation to be represented by the tree at the bottom, $T$. Second, we added information sets at terminal nodes. The reason is that in order to discuss notions of equilibrium under unawareness, it will be useful to analyze also the players' views at the end of the game. The information sets also model interactive reasoning of players. For instance, at his information set in tree $\bar{T}$, player 2 knows that initially player 1 is unaware of action $m$ and views the game as given by the lower tree $T$. Moreover, he knows that player 1 considers player 2's state of mind be given by the information set in $T$ once she takes action $\ell_1$. To complete the description, note that players receive a payoff at each terminal node. The first component at each terminal node refers to player 1's payoff whereas the second component refers to player 2's payoff.

\begin{figure}[h!]
\caption{Game of Example 1 after being played once}
\label{example1b}
\begin{center}
\includegraphics[scale=0.35]{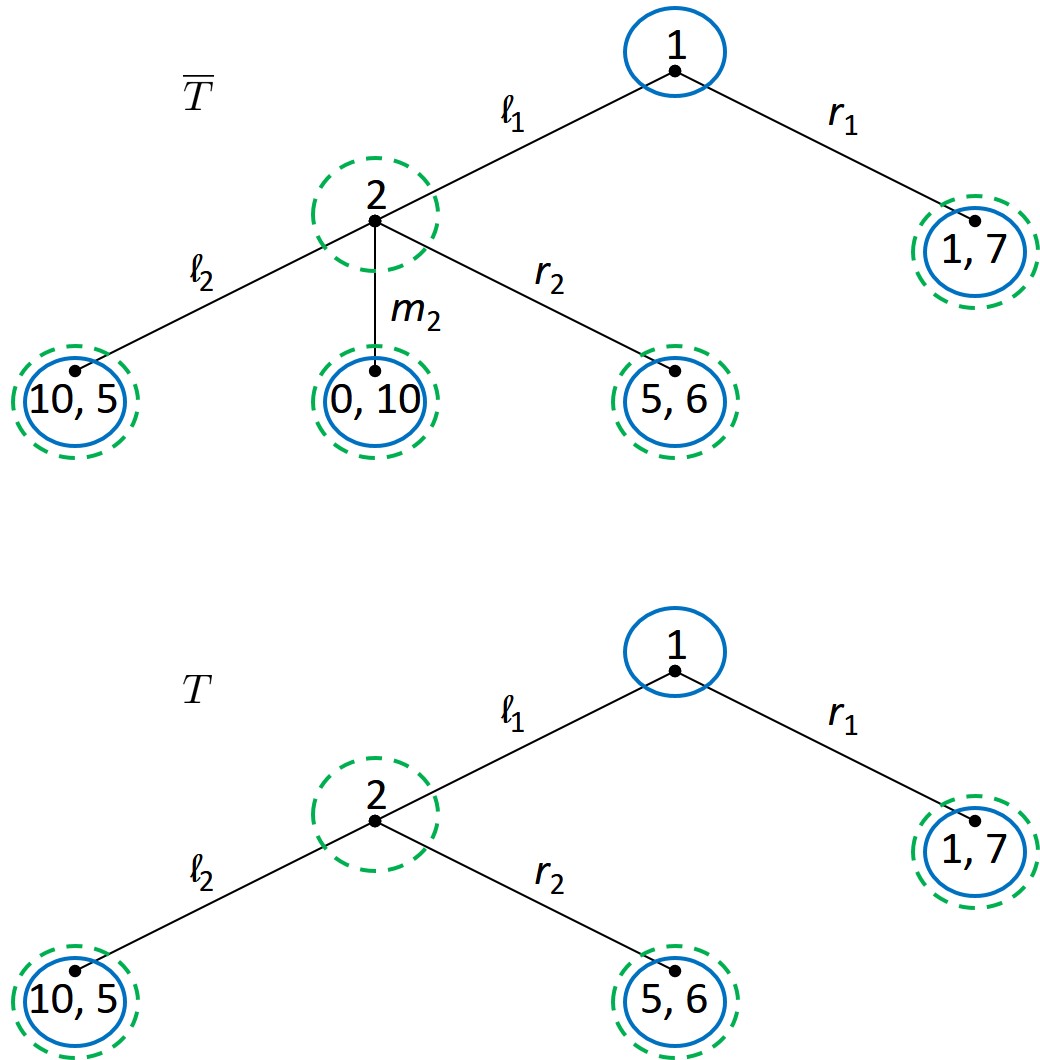}
\end{center}
\end{figure}

What is equilibrium in this game? A basic requirement is that in equilibrium players should play rational. That is, each player at each information set where (s)he is to move should play an action that maximizes her expected payoff w.r.t. her belief over the opponent's behavior. At the beginning of the game, player 1 thinks that she faces the situation depicted in lower tree $T$. Clearly, with this mindset only action $\ell_1$ is rational because no matter what she expects player 2 to do, she obtains a higher expected payoff from playing $\ell_1$ than $r_1$. At the information set in the upper tree $\bar{T}$, player 2 is aware of his action $m_2$. Since $m_2$ strictly dominates any of his other actions, the only rational action for player 2 at this information set is to choose $m_2$. Thus, the path of play emerging from rational play is $(\ell_1, m_2)$ with player 1 obtaining zero payoff and player 2 obtaining a payoff of $10$. The game is on purpose trivial so that all solution concepts to games with unawareness that have been proposed in the literature (Halpern and Rego, 2014, Rego and Halpern, 2012, Feinberg, 2020, Li 2008, Grant and Quiggin, 2013, Ozbay, 2007, Heifetz, Meier, and Schipper, 2013, 2021) yield the same profile of strategies. Yet, we strongly believe that this profile of rational strategies cannot reasonably be called an equilibrium in this setting because any profile of strategies in which player 1 chooses $\ell_1$ and player 2 chooses $m_2$ is impossible to interpret as a steady-state of a learning process. After players choose rationally in the game, player 1's awareness has changed. She discovered action $m_2$ of player 2. This is symbolized by player 1's information set at the terminal node after $m_2$ in the tree $\bar{T}$. Thus, the ``next'' time, players do not play the game of Figure~\ref{example1a} but a ``discovered version'' of it in which player 1 is aware of action $m_2$ upfront. This discovered game is depicted in Figure~\ref{example1b}. At the beginning of the game, player 1's information set is now in the upper tree $\bar{T}$. Consequently she is aware of all actions of all players. She will not be surprised by any terminal node as her information sets at terminal nodes in the upper tree $\bar{T}$ also lie in this tree. The lower tree $T$ becomes in some sense redundant as players are now commonly aware of the strategic situation modeled by the upper $\bar{T}$. Yet, since they are aware, they can envision themselves also in a situation in which both players are unaware of $m_2$, which is what now $T$ represents although this counterfactual mindset is not behaviorally relevant. The games in Figure~\ref{example1a} and~\ref{example1b} differ only in the information sets. The information sets of the game of Figure~\ref{example1a} are updated such that information is preserved and just the awareness gained from play of the game in Figure~\ref{example1a} is reflected in the updated information sets of the game in Figure~\ref{example1b}.

\begin{figure}[h!]
\caption{Discovery process in Example 1\label{example1_process}}
\begin{center}
\includegraphics[scale=0.5]{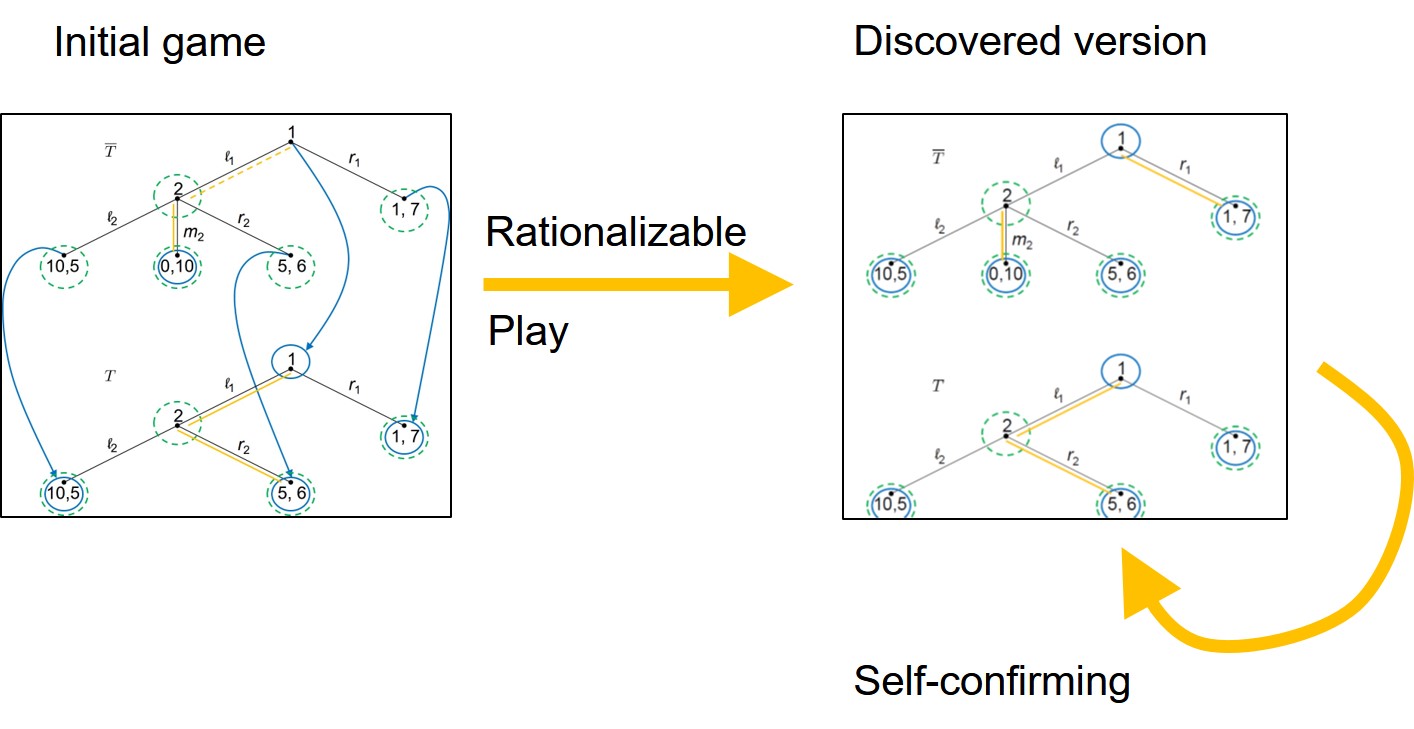}
\end{center}
\end{figure}

In the discovered version shown in Figure~\ref{example1b}, the only rationalizable action for player 1 at the beginning of the game is to choose $r_1$ in $\bar{T}$. Nothing can be discovered anymore. The game in Figure~\ref{example1b} becomes an absorbing state of the discovering process. Any steady-state of a learning and discovery process must prescribe $r_1$ for player 1 in $\bar{T}$. The discovery process is schematically depicted in Figure~\ref{example1_process}. There are two states, left is the game of Figure~\ref{example1a} and right the game of Figure~\ref{example1b}. The transition is via the rational profile of strategies $(\ell_1, m_2)$ in the initial game. Once the right game is reached, it is absorbing.

To sum up, we note first that games with unawareness may not possess solutions that can be interpreted as steady-states of a learning process (see the game in Figure~\ref{example1a}). Second, an equilibrium notion capturing the idea of a steady-state of a learning and discovery process in games with unawareness must not only involve the usual conditions on behavior of players but must also impose restrictions on their views of the strategic situation. That is, their representations of the strategic situation must be consistent with their behavior and behavior must be consistent with their representations of the strategic situations. Implicitly, the process of discovering actions must have reached a steady-state as well. To emphasize this, we will use the terminology of self-confirming game.\footnote{At a first glance, this terminology may sound odd because in standard game theory, the representation of the strategic situation is given and players' behavior is endogenous. But the point of our terminology is precisely that in our setting the representation of the strategic situation becomes endogenous too.} The game of Figure~\ref{example1a} is not a self-confirming game while the game of Figure~\ref{example1b} is. When players play the game in Figure~\ref{example1b}, no further changes of awareness are feasible. The representation of the game (together with rationality) and player 1's belief in player 2's rationality induces the behavior and what is observed with this behavior just confirms the representation. In contrast, when players play rationally in the game depicted in Figure~\ref{example1a}, then player 1 discovers features of the game that she was previously unaware of. That is, player 1's initial representation of the game is destroyed and a new version is discovered in which optimal behavior differs from optimal behavior in the initial version.\\

\begin{figure}[h!]
\caption{(Initial) Game of Example 2\label{example2a}}
\begin{center}
\includegraphics[scale=0.3]{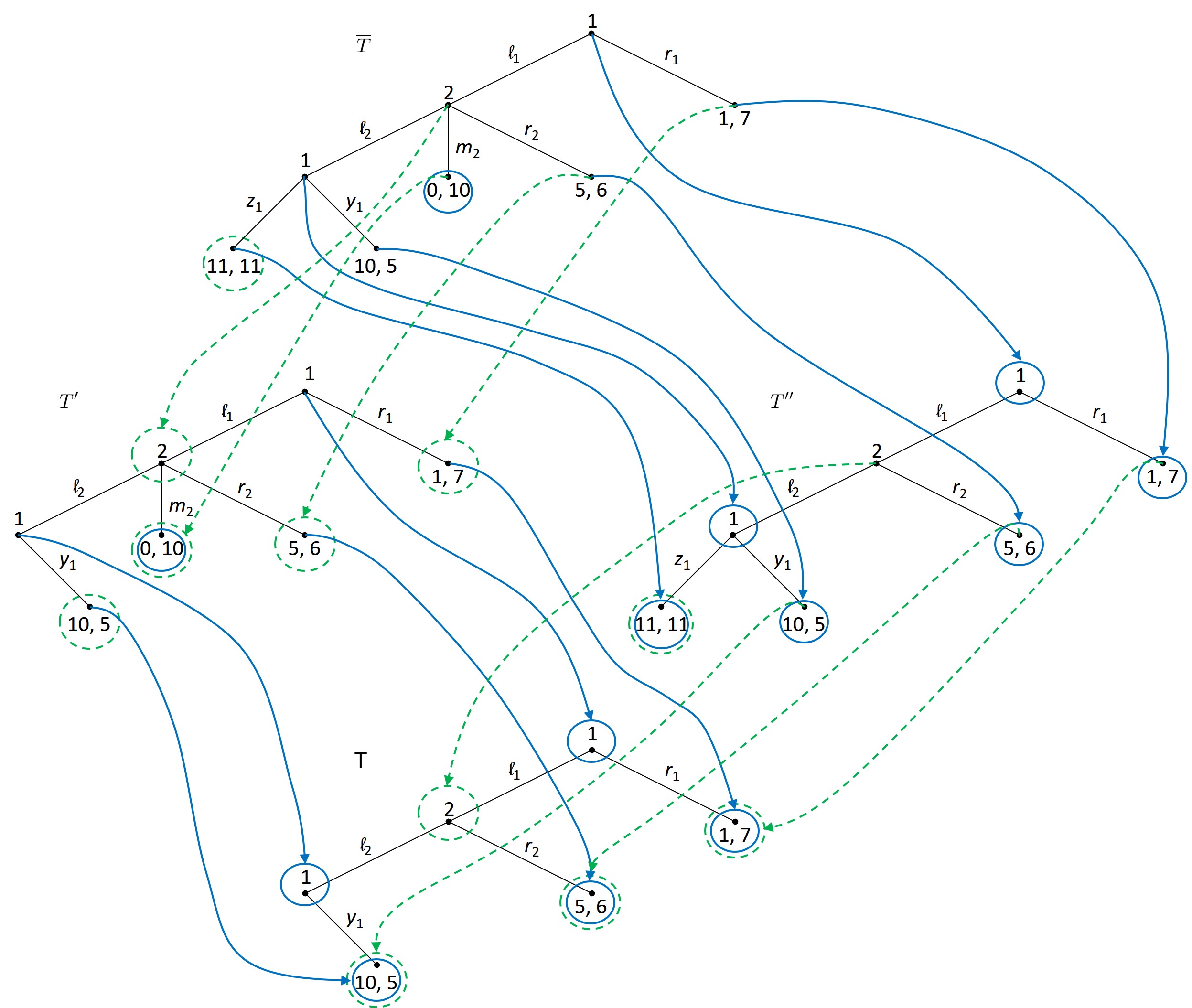}
\end{center}
\end{figure}

\noindent \textbf{Example 2 } Example 1 should not mislead the reader to believe that self-confirming games must involve common awareness of the strategic situation and that rational discovery would justify restricting the focus to standard games like given by the upper tree $\bar{T}$ in Figure~\ref{example1b}. One can easily extend the example to discuss a situation in which the self-confirming game involves understandings of the strategic situation that differ among players. For instance, Figure~\ref{example2a} depicts a slightly more enriched version of the prior example in which initially each player is aware of an action of which the other player is unaware. Note first that trees $T'$ and $T$ together with their information sets are just as in Figure~\ref{example1a} except except that latter collapses action $y_1$. Trees $\bar{T}$ and $T''$ are similar but contain an additional action $z_1$ for player 1 after player 2 takes action $\ell_2$. Initially, player 1 is aware of action $z_1$ but unaware of action $m_2$. This is indicated by the blue arrow that leads from the initial node in tree $\bar{T}$ to the blue information set containing the initial node of tree $T''$. In contrast, player 2 is initially unaware of action $z_1$ but aware of his action $m_2$. This is shown by green intermitted arrows from his nodes after history $\ell_1$ or $r_1$ in tree $\bar{T}$ to the green intermitted information sets containing the analogous node in tree $T'$.
\begin{figure}[h!]
\caption{Game of Example 2 after being rationally played once\label{example2b}}
\begin{center}
\includegraphics[scale=0.3]{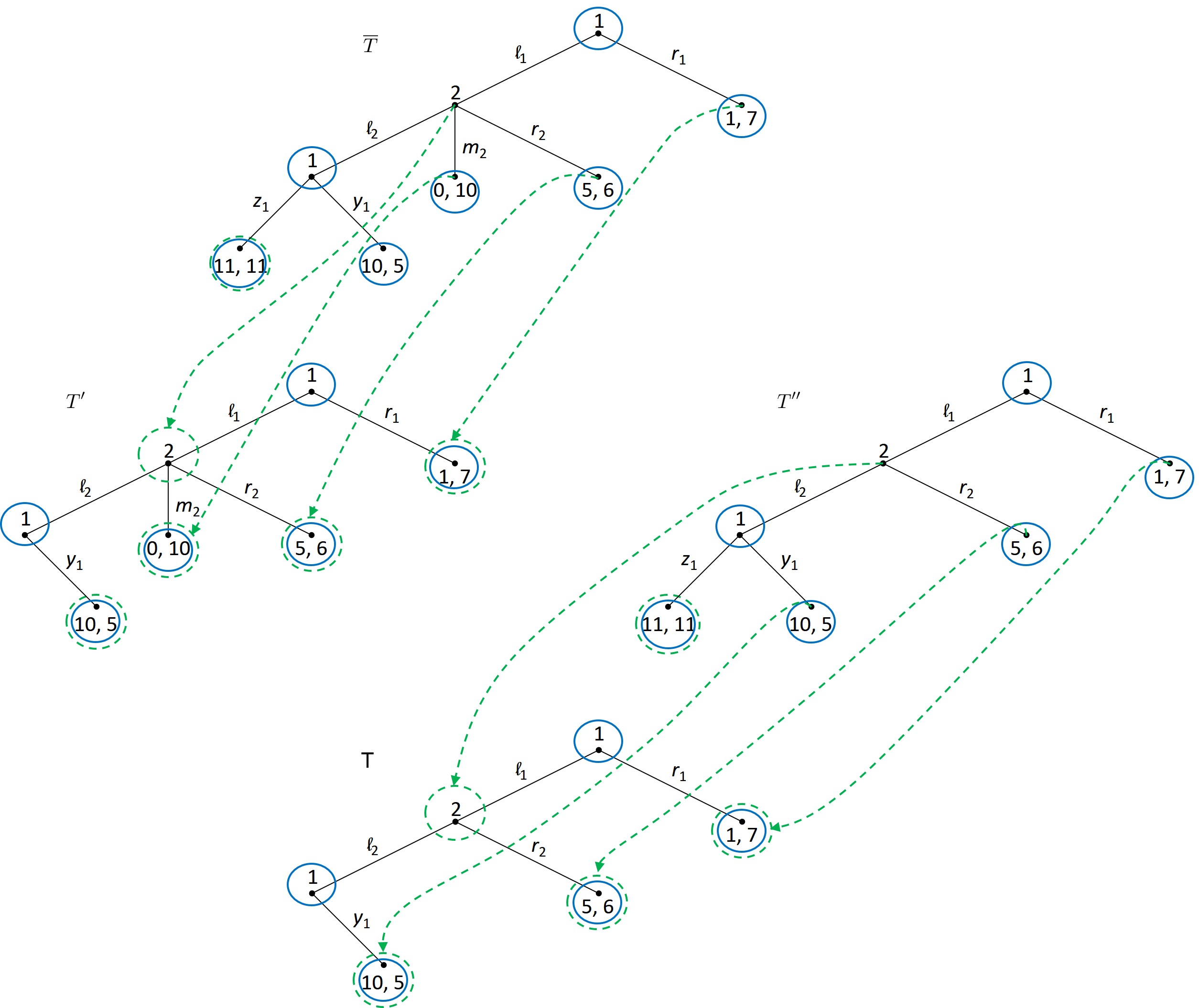}
\end{center}
\end{figure}

It is easy to see that for player 1, action $z_1$ strictly dominates action $y_1$ conditional on reaching player $1$'s second information set in tree $T''$. Yet, player 2 will not play $\ell_2$ since he is unaware of $z_1$ and the expected payoff from playing $m_2$ is strictly larger than the expected payoff from playing $\ell_2$. In fact, to him action $m_2$ strictly dominates $\ell_2$. Consequently, player 2 will not be able to discover action $z_1$ and will remain unaware of it. Together with arguments about optimal play in Example 1, it implies that after the game is optimally played once by both players, the representation must change to the one depicted in Figure~\ref{example2b}. In this discovered version, player 1 is now aware of action $m_2$ in addition to her action $z_1$ (i.e., she ``lives'' in tree $\bar{T}$). This is indicated by the blue information sets in the upmost tree $\bar{T}$, which are now different from Figure~\ref{example2a}. Player 1 realizes that player 2 remains unaware of action $z_1$, and believes that player 2 views the strategic situation as represented by $T'$ and $T$. Optimal play is as in the game of Figure~\ref{example1b}. Thus, player 2 will not become aware of actions $z_1$ and differences in players' awareness persist. The game of Figure~\ref{example2b} is self-confirming since it is an absorbing state of the discovery process in which in every game played along the discovery process all players play rationalizable strategies.
\begin{figure}[h!]
\caption{Non-rationalizable Discovered Version of the Game in Example 2\label{example2c}}
\begin{center}
\includegraphics[scale=0.3]{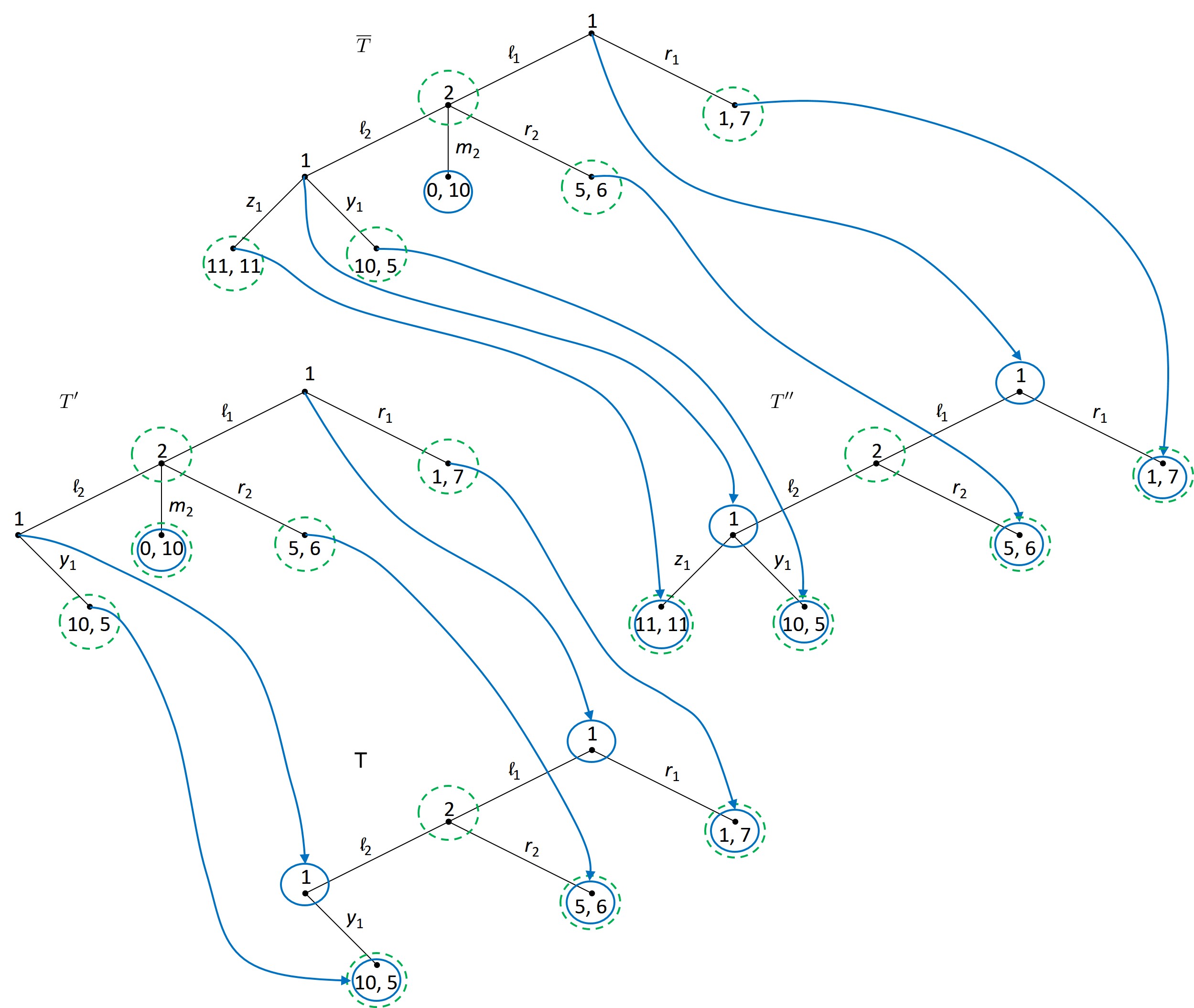}
\end{center}
\end{figure}
\begin{figure}[h!]
\caption{Non-rationalizable Self-confirming game in Example 2\label{example2d}}
\begin{center}
\includegraphics[scale=0.3]{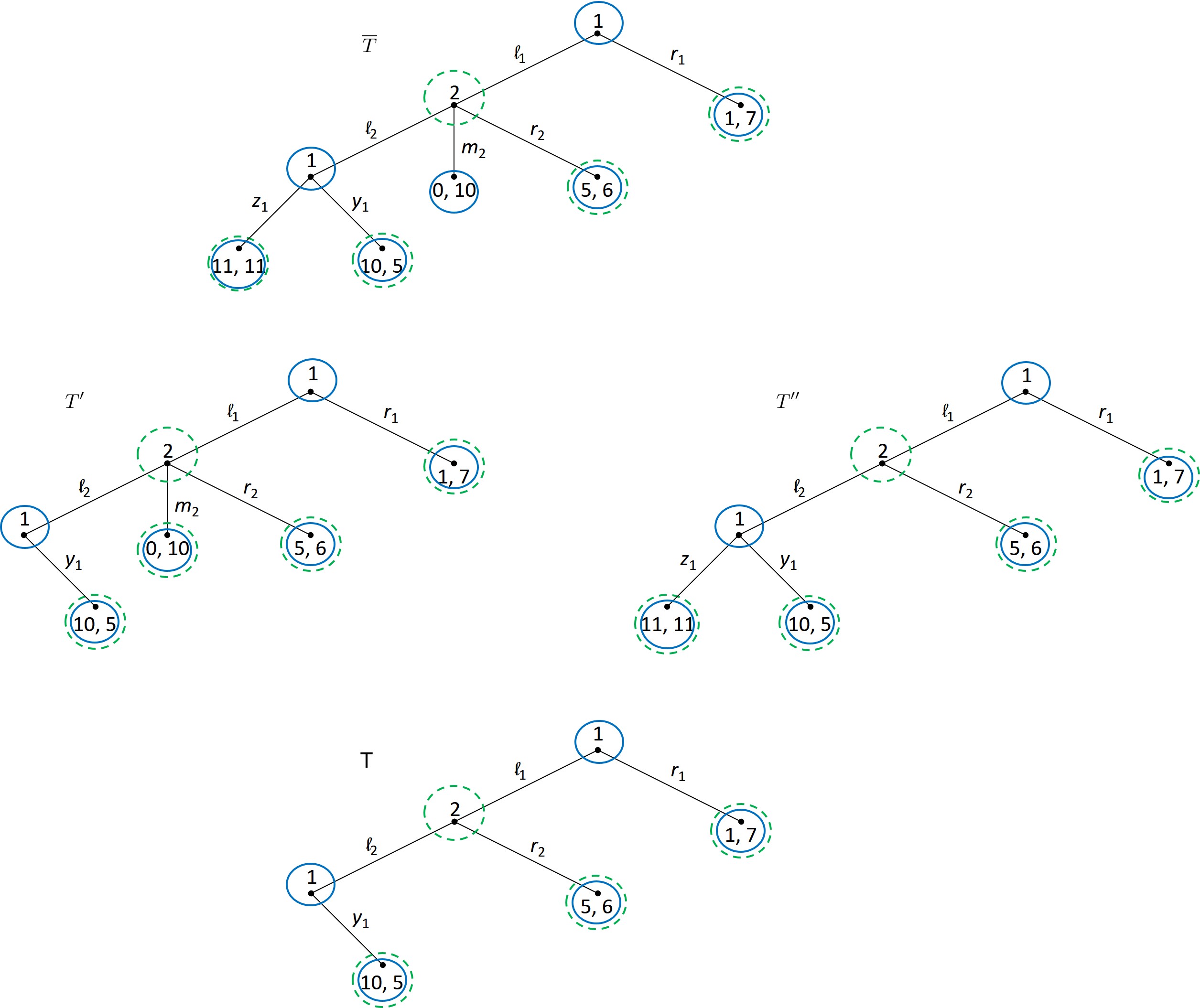}
\end{center}
\end{figure}

The fact that some player may remain unaware of some actions is probably symptomatic for most strategic situations in reality. Players interact with different views of the game and settle in a steady-state of behavior that does not allow them to learn or, more precisely, to discover that they have different views. Of course, in the situation modelled above, player 1 has an incentive to raise player 2's awareness of action $z_1$ through some form of communication because it leads to the payoff dominant terminal history. Yet, the current game form does not allow for such communication. Moreover, one can easily come up with other examples where no player has an incentive to raise the other player's awareness of an action (e.g., a strictly dominated action) and unawareness persists.
\begin{figure}[h!]
\caption{Discovery processes in Example 2\label{example2_process}}
\begin{center}
\includegraphics[scale=0.28]{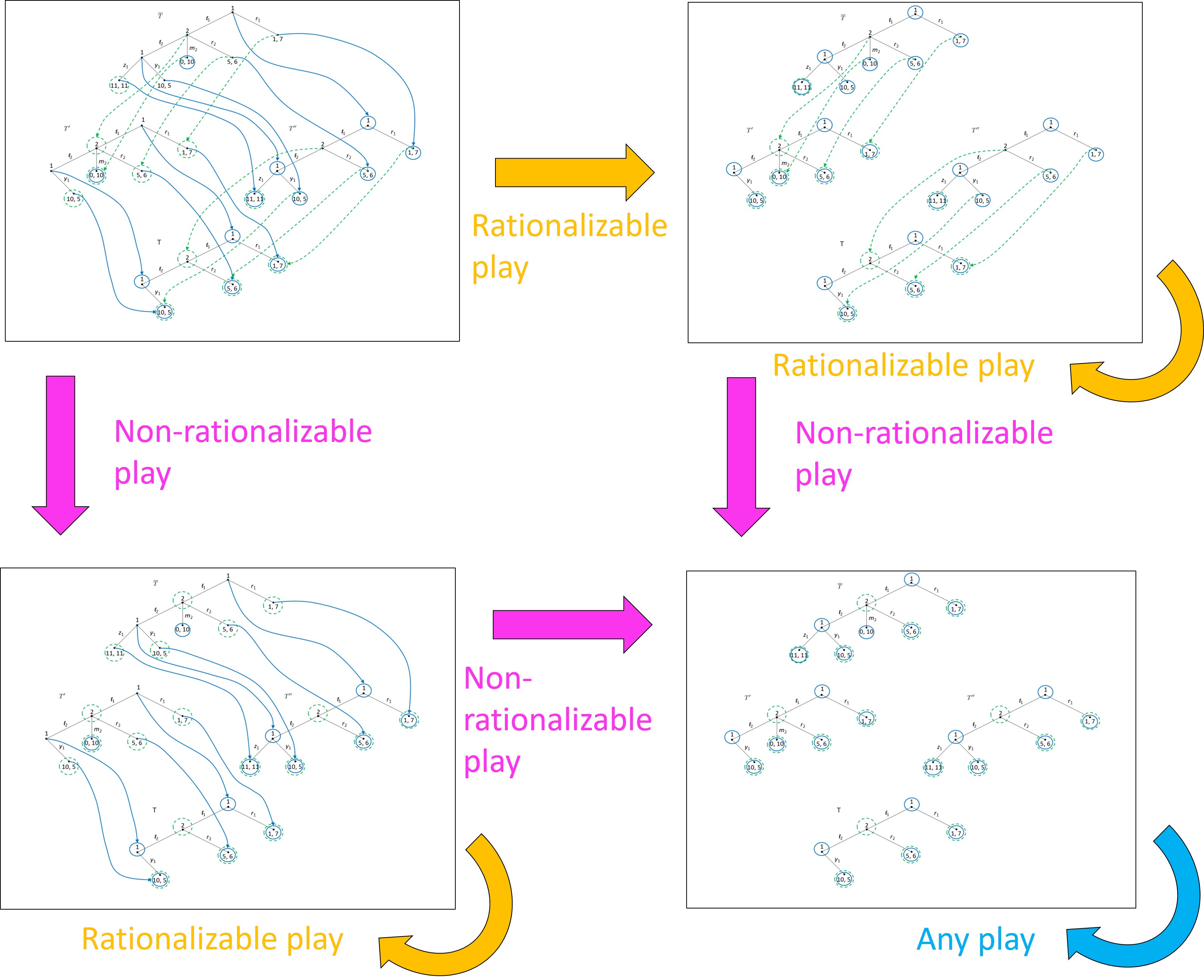}
\end{center}
\end{figure}

The game in Figure~\ref{example2b} is not just a self-confirming game. It is a rationalizable self-confirming game because it is a discovered version of the game in Figure~\ref{example2a} \emph{after players played rationalizable strategies}. If players play differently, they may discover other versions. For instance, the game in Figure~\ref{example2c} is a discovered version of the game in Figure~\ref{example2a} after player 2 played $\ell_2$ and player 1 played $z_1$. Since it is not rational for player 2 to play $\ell_2$ in tree $T'$ of the game in Figure~\ref{example2a}, this discovered version in Figure~\ref{example2c} is not rationalizable. Hence, it cannot be a rationalizable self-confirming game. If in the game of Figure~\ref{example2c} players play rationally, then player 2 never plays $m_2$. Hence, player 1 remains unaware of $m_1$ and the game is self-confirming of the discovery process that allows for non-rationalizable strategies in Figure~\ref{example2a} and restricting to rationalizable strategies in Figure~\ref{example2c}. Yet, the game of Figure~\ref{example2c} is not a self-confirming game when allowing for non-rationalizable strategies also in Figure~\ref{example2c}. If player 2 plays $m_2$ in the game of Figure~\ref{example2c}, then player 1 becomes aware of it. The discovered version is the game depicted in Figure~\ref{example2d}. Note that player 2 playing $m_2$ is non-rationalizable in the game of Figure~\ref{example2c} since he is already aware of the payoff dominant outcome emerging from history $(\ell_1, \ell_2, z_1)$. The game of Figure~\ref{example2d} can also be reached via non-rationalizable play of player 2 in the game of Figure~\ref{example2b}. Moreover, no matter what is played in the game of Figure~\ref{example2d}, no players view is changed. Thus, Figure~\ref{example2d} is the self-confirming game when any pure strategy is allowed.

Figure~\ref{example2_process} summarizes the discovery processes. From the initial state in the upper left corner, rationalizable play yields the self-confirming game in the upper right corner. Non-rationalizable play eventually leads to the self-confirming game in the lower right corner. The discussion so far demonstrates four points: First, non-rationalizable discoveries may lead to views that are different from views emerging from from rationalizable discoveries. Second, rationalizability may select among discovery processes and hence among views that may emerge. Third, full common awareness may not emerge among players with rationalizable discovery processes. Fourth, the fact that rationalizable discovery processes lead to different views from non-rationalizable discovery processes and that full common awareness may not emerge with rationalizable discovery processes is both relevant for limit behavior and outcomes.

\section{Extensive-Form Games with Unawareness\label{model}}

In this section, we outline extensive-form game with unawareness as introduced by Heifetz, Meier, and Schipper (2013) together with some crucial extensions especially required for our analysis. As with any work that requires spelling out extensive-form games, some notation is unavoidable.

To define an extensive-form game with unawareness $\Gamma$, consider first, as a building block, a finite game with perfect information and possibly simultaneous moves. The major purpose is just to outline all physical moves. There is a finite set of players $I$ and possibly a special player ``nature'' with index $0$. We denote by $I^0$ the set of players including nature. Further, there is a nonempty finite set of ``decision'' nodes $\bar{D}$ and a player correspondence $P: \bar{D} \longrightarrow 2^{I^0} \setminus \{\emptyset\}$ that assigns to each node $n \in \bar{D}$, a nonempty set of ``active'' players $P(n) \subseteq I^0$. (That is, we allow for simultaneous moves.) For every decision node $n \in \bar{D}$ and player $i \in P(n)$ who moves at that decision node, there is a nonempty finite set of actions $A_{n}^{i}$. Moreover, there is a set of terminal nodes $\bar{Z}$. Since we will also associate information sets with terminal nodes for each player, it will be useful to extent $P$ to $\bar{Z}$ by $P(z) = I$ and let $A_z^i \equiv \emptyset$ for all $i \in I$, $z \in \bar{Z}$. Finally, each terminal node $z \in \bar{Z}$ is associated with a vector of payoffs $(u_i(z))_{i \in I}$. We require that nodes in $\bar{N} := \bar{D} \cup \bar{Z}$ constitute a tree denoted by $\bar{T}$. That is, nodes in $\bar{N}$ are partially ordered by a precedence relation $\lessdot$ with which $(\bar{N}, \lessdot)$ forms an arborescence (that is, the predecessors of each node in $\bar{N}$ are totally ordered by $\lessdot$). There is a unique node in $\bar{N}$ with no predecessors (i.e., the root of the tree). For each decision node $n \in \bar{D}$ there is a bijection $\psi_n$ between the action profiles $\prod_{i \in P(n)} A_{n}^{i}$ at $n$ and $n$'s immediate successors. Finally, any terminal node in $\bar{Z}$ has no successors.

Note that so far we treat nature like any other player except that at terminal nodes we do not assign payoffs to nature.\footnote{Alternatively, we could assign at every terminal node the same payoff to nature.} We do not need to require that nature moves first or that nature moves according to a pre-specified probability distribution (although these assumptions can be imposed in our framework). Nature may also move simultaneously with other players.

Consider now a join-semilattice $\mathbf{T}$ of subtrees of $\bar{T}$.\footnote{A join semi-lattice is a partially ordered set in which each pair of elements has a join, i.e., a least upper bound.} A subtree is defined by a subset of nodes $N \subseteq \bar{N}$ for which $(N, \lessdot)$ is also a tree. Two subtrees $T', T'' \in \mathbf{T}$ are ordered, written
\begin{equation*}
T' \preceq T''
\end{equation*}
if the nodes of $T'$ constitute a subset of the nodes of $T''$.

We require three properties:
\begin{enumerate}
\item All the terminal nodes in each tree $T \in \mathbf{T}$ are in $\bar{Z}$. That is, we do not create ``new'' terminal nodes.

\item For every tree $T \in \mathbf{T}$, every node $n \in T$, and every active player $i \in P(n)$ there exists a nonempty subset of actions $A_{n}^{i,T} \subseteq A_{n}^{i}$ such that $\psi_{n}$ maps the action profiles $A_{n}^{T} = \prod_{i \in P(n)} A_{n}^{i,T}$ bijectively onto $n$'s successors in $T$. We say that at node $n$ action profile $a_n \in A_n^T$ leads to node $n'$ if $\psi_n(a_n) = n'$.

\item For any tree $T \in \mathbf{T}$, if for two decision nodes $n, n' \in T$ with $i \in P(n) \cap P(n')$ it is the case that $A_{n}^{i} \cap A_{n'}^{i} \neq \emptyset $, then $A_{n}^{i} = A_{n'}^{i}$.
\end{enumerate}

Within the family $\mathbf{T}$ of subtrees of $\bar{T}$, some nodes $n$ appear in several trees $T \in \mathbf{T}$. In what follows, we will need to designate explicitly appearances of such nodes $n$ in different trees as distinct objects. To this effect, in each tree $T \in \mathbf{T}$ label by $n_{T}$ the copy in $T$ of the node $n \in \bar{N}$ whenever the copy of $n$ is part of the tree $T$, with the requirement that if the profile of actions $a_{n} \in A_{n}$ leads from $n$ to $n'$ in $\bar{T}$, then $a_{n_T}$ leads also from the copy $n_{T}$ to the copy $n_{T}'$. More generally, for any $T, T', T'' \in \mathbf{T}$ with $T \preceq T' \preceq T''$ such that $n \in T''$, $n_{T'}$ is the copy of $n$ in the tree $T'$, $n_T$ is the copy of $n$ in the tree $T$, and $(n_{T'})_{T}$ is the copy of $n_{T'}$ in the tree $T$, we require that ``nodes commute'', $n_T = (n_{T'})_T$. For any $T \in \mathbf{T}$ and any $n \in T$, we let $n_T := n$ (i.e., the copy of $n \in T$ in $T$ is $n$ itself).

Denote by $\mathbf{D}$ the union of all decision nodes in all trees $T \in \mathbf{T}$, by $\mathbf{Z}$ the union of terminal nodes in all trees $T \in \mathbf{T}$, and by $\mathbf{N} = \mathbf{D} \cup \mathbf{Z}$. Copies $n_{T}$ of a given node $n$ in different subtrees $T$ are now treated distinct from one another, so that $\mathbf{N}$ is a disjoint union of sets of nodes.\footnote{Bold capital letters refer to sets of elements across trees.}

In what follows, when referring to a node in $\mathbf{N}$ we will typically avoid the subscript indicating the tree $T$ for which $n \in T$ when no confusion arises. For a node $n \in \mathbf{N}$ we denote by $T_{n}$ the tree containing $n$.

Denote by $N^{T}$ the set of nodes in the tree $T \in \mathbf{T}$. Similarly, for any $i \in I^0$ denote by $D_i^T$ the set of decision nodes in which player $i$ is active in the tree $T \in \mathbf{T}$. Finally, denote by $Z^T$ the set of terminal nodes in the tree $T \in \mathbf{T}$.

For each node $n \in \mathbf{N}$ (including terminal nodes in $\mathbf{Z}$), define for each active player $i \in P(n) \setminus \{0\}$ a nonempty information set $h_i(n)$. These information sets shall satisfy properties U0, U1, U4, U5, and I2-I6 introduced by Heifetz, Meier, and Schipper (2013). These properties generalize properties of standard extensive-form games to extensive-form games with unawareness and impose properties of unawareness (paralleling properties of static unawareness structures in Heifetz, Meier, and Schipper, 2006). While we refer to Heifetz, Meier, and Schipper (2013) and Schipper (2018) for further discussions and graphical illustrations of these properties, we do list these properties at the beginning of the proof of Proposition~\ref{well-defined} in the Appendix.

Information sets model both information and awareness. Let $T_n$ denote the tree that contains node $n$. At a node $n$ of the tree $T_{n} \in \mathbf{T}$, the player may conceive the feasible paths to be described by a different (i.e., less expressive) tree $T' \in \mathbf{T}$, $T' \preceq T_n$. In such a case, her information set will be a subset of $T'$ rather than $T_{n}$ and $n$ will not be contained in the player's information set at $n$. An example is the initial information set of player 1 in Figure~\ref{example1a}.

In order to define a notion of self-confirming equilibrium we also need to consider the players' views at terminal nodes. Thus, we devise also information sets of terminal nodes that model both the players' information and awareness at the ends of the game. This is different from Heifetz, Meier, and Schipper (2013) but akin to signal, outcome, or feedback functions in some works on self-confirming equilibrium, see for instance Battigalli and Guaitoli (1997), Battigalli et al. (2015), and Battigalli and Bordolli (2020). The following property is imposed on information set of terminal nodes only:

\begin{itemize}

\item[I7] Information sets consistent with own payoff information: If $h_i(z) \subseteq T$ then $h_i(z) \subseteq Z^{T}$. Moreover, if $z' \in h_i(z)$ then $u_i(z') = u_i(z)$.

\end{itemize}

At any terminal node, a player considers only terminal nodes. That is, she knows that the game ended. Moreover, any two terminal nodes that a player cannot distinguish must yield her the same payoff because otherwise she could use her payoffs to distinguish among these terminal nodes. This implies that at the end of the game each player knows her own payoff. Note that this assumption does not rule out imperfect observability of \emph{opponents'} payoffs. It also does not rule out that the player may not perfectly observe the terminal node.

\begin{figure}[h!]
\caption{Property I7\label{properties}}
\begin{center}
\includegraphics[scale=0.25]{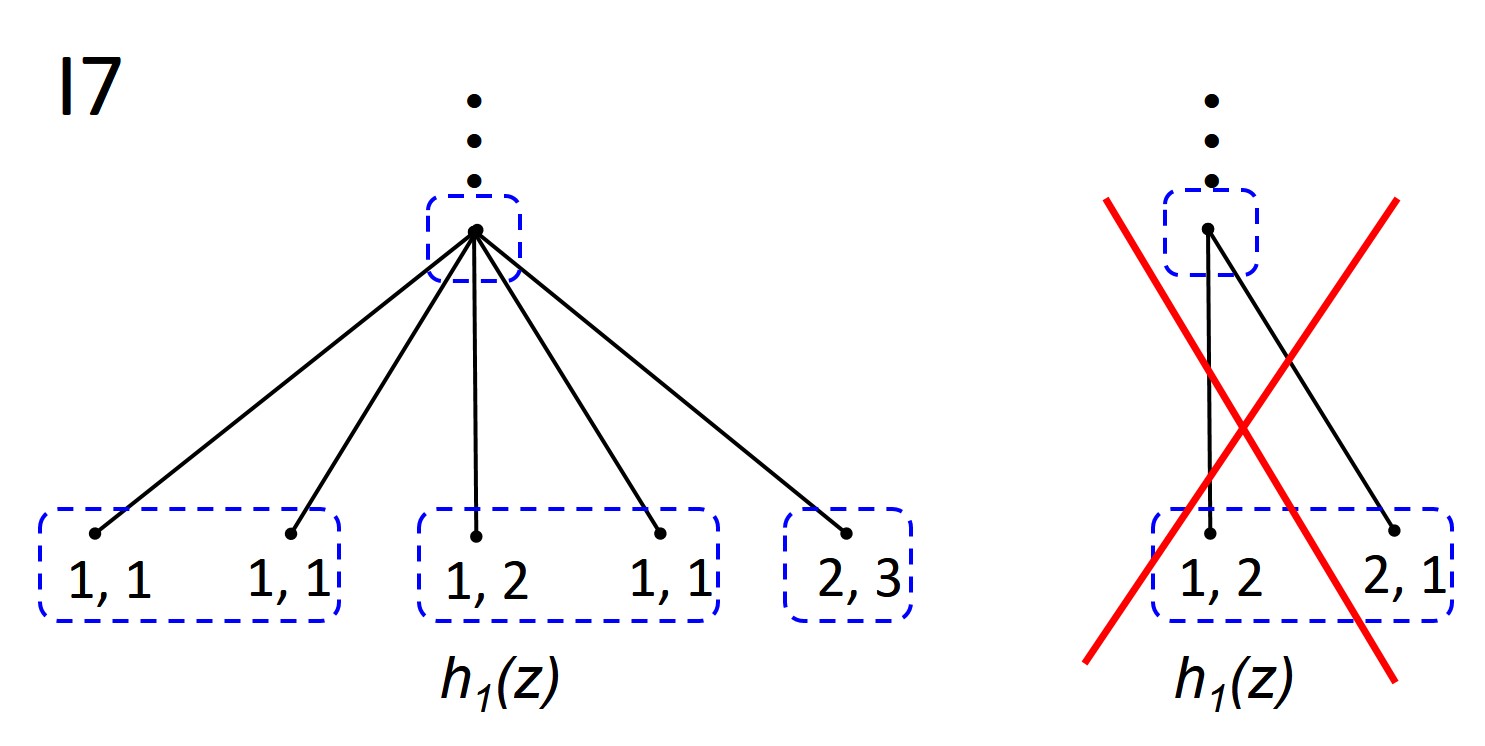}
\end{center}
\end{figure}

Figure~\ref{properties} illustrates property I7. For this illustration, assume that the player moving at the node that is immediately preceding the terminal nodes is the player whose payoffs are indicated by the first component of the payoff vectors that are attached to the terminal nodes.

We denote by $H_{i}$ the set of $i$'s information sets in all trees. For an information set $h_{i} \in H_{i}$, we denote by $T_{h_{i}}$ the tree containing $h_{i}$. For two information sets $h_{i}, h_{i}^{\prime}$ in a given tree $T,$ we say that $h_{i}$ precedes $h_{i}^{\prime }$ (or that $h_{i}^{\prime }$ succeeds $h_{i}$) if for every $n^{\prime} \in h_{i}^{\prime }$ there is a path $n,...,n^{\prime }$ in $T$ such that $n \in h_{i}$. We denote it by $h_{i}\rightsquigarrow h_{i}^{\prime }$. The perfect recall property I6 guarantees that with the precedence relation $\rightsquigarrow $ player $i$'s information sets $H_{i}$ form an arborescence: For every information set $h_{i}^{\prime }\in H_{i}$, the information sets preceding it $\left\{ h_{i} \in H_{i}: h_{i} \rightsquigarrow h_{i}^{\prime }\right\}$ are totally ordered by $\rightsquigarrow$. Note also that if $n \in h_i$ we can write $A_{h_i}$ for $A^i_{n}$ (see Heifetz, Meier, and Schipper, 2013).

For trees $T, T^{\prime } \in \mathbf{T}$ we denote $T \rightarrowtail T^{\prime }$ whenever for some node $n\in T$ and some player $i\in P(n)$ it is the case that $h_{i}(n) \subseteq T'$. Denote by $\hookrightarrow $\ the transitive closure of $\rightarrowtail$. That is, $T \hookrightarrow T^{\prime \prime}$ if and only if there is a sequence of trees $T, T^{\prime }, \dots, T^{\prime \prime } \in \mathbf{T}$ satisfying $T \rightarrowtail T^{\prime} \rightarrowtail \dots \rightarrowtail T^{\prime\prime}$. For instance, in Figure~\ref{example2a} we have $\bar{T} \rightarrowtail T'$ and $T' \rightarrowtail T$ as well as $\bar{T} \rightarrowtail T''$ and $T'' \rightarrowtail T$. Clearly, $\bar{T} \hookrightarrow T$.

An \emph{extensive-form game with unawareness} $\Gamma $ consists of a join-semilattice $\mathbf{T}$ of subtrees of a tree $\bar{T}$ satisfying properties 1--3 above, along with information sets $h_i(n)$ for every $n \in T$ with $T \in \mathbf{T}$ and $i \in P(n)$, and payoffs satisfying properties U0, U1, U4, U5, and I2-I7 (see Appendix).

For every tree $T \in \mathbf{T}$, the $T$\emph{-partial game} is the join-semisublattice of trees including $T$ and all trees $T^{\prime}$ in $\Gamma$ satisfying $T \hookrightarrow T^{\prime}$, with information sets as defined in $\Gamma$. A $T$-partial game is a extensive-form game with unawareness, i.e., it satisfies all properties 1--3, U0, U1, U4, U5, and I2-I7 (see Appendix). For instance, in Figure~\ref{example2a} the sublattice $\{T', T\}$ together with all information sets in those trees forms the $T'$-partial game. In fact, modulo collapsing of actions, it is the game with unawareness of Figure~\ref{example1a}.

We denote by $H_{i}^{T}$ the set of $i$'s information sets in the $T$-partial game, $T \in \mathbf{T}$. This set contains not only $i$'s information sets in the tree $T$ but also in all trees $T' \in \mathbf{T}$ with $T \hookrightarrow T'$.

Further, we denote by $H_i^{\mathbf{D}}$ ($H_i^{T, \mathbf{D}}$, resp.) the set of $i$'s information sets of decision nodes (in the $T$-partial game, resp.) and by $H_i^{\mathbf{Z}}$ ($H_i^{T, \mathbf{Z}}$, resp.) the set of $i$'s information sets of terminal nodes (in the $T$-partial game, resp.).

\subsection{Strategies}

For any collection of sets $(X_i)_{i \in I^0}$ we denote by
$$X := \prod_{i \in I^0} X_{i}, \quad \mbox{ and } \quad X_{-i} := \prod_{j \in I^0 \setminus \{i\}} X_{j}$$ with typical elements $x$ and $x_{-i}$, respectively. For any collection of sets $(X_i)_{i \in I^0}$ and any tree $T \in \mathbf{T}$, we denote by $X_i^T$ the set of objects in $X_i$ restricted to the tree $T$ and analogously for $X^T$ and $X^T_{-i}$, where ``restricted to the tree $T$'' will become clear from the definitions below.

A \emph{pure strategy} for player $i \in I$ in game $\Gamma$,
\begin{equation*}
s_{i} \in S_{i} := \prod_{h_{i}\in H_{i}^{\mathbf{D}}} A_{h_{i}}
\end{equation*}
specifies an action of player $i$ at each of her information sets $h_{i}\in H_{i}^{\mathbf{D}}$ of decision nodes. We let
\begin{equation*}
s_0 \in S_0 := \prod_{n \in \mathbf{D}_0} A^0_{n}
\end{equation*} denote the ``strategy'' of nature, with $D_0$ denoting the ``decision'' nodes of nature.

For any player $i \in I$, strategy $s_{i}$, and node $n \in D_{i}^{T_n}$, player $i$'s action at $n$ is $s_i(h_{i}(n))$. Thus, by U1 and I4 the strategy $s_{i}$ specifies what player $i \in I$ does at each of her active nodes $n \in D_{i}^{T_n}$, both in the case that $n \in h_i(n)$ \emph{and} in the case that $h_i(n) $ is a subset of nodes of a tree which is distinct from the tree $T_{n}$ to which $n$ belongs. In the first case, when $n \in h_i(n)$, we can interpret $s_i(h_i(n))$ as the action chosen by player $i$ in node $i$. In the second case, when $n \notin h_i(n)$, $s_i(h_i(n))$ cannot be interpreted as the action chosen ``consciously'' by player $i$ in $n$ since he is not even aware of $T_n$. Instead, his state of mind at $n$ is given by his information set $h_i(n)$ in a tree less expressive than $T_n$ (denoted by $T_{h_i}$). Thus, $s_i(h_i(n))$ is the physical move of player $i$ at $n$ in tree $T_n$ induced by his ``consciously'' chosen action at his information set $h_i(n)$ in tree $T_{h_i(n)}$ (with $T_n \succ T_{h_i(n)}$ by U0). As an example, consider player 1 in the game of Figure~\ref{example1a}. At his first decision node in the upper tree $\bar{T}$, the root of the tree, player 1's information set consists of the corresponding node in the less expressive tree $T$. The optimal strategy of player 1 may assign $\ell_1$ to his information set in the lower tree $T$. But it also induces action $\ell_1$ at the root of the upper tree $\bar{T}$. For a further discussion of the interpretation of strategies in extensive-form games with unawareness, see Heifetz, Meier, and Schipper (2013). A strategy of a player becomes meaningful as an object of beliefs of other players. How ``much'' of a player's strategy other players can conceive depend on their awareness given by the tree in which their information set is located. This leads to the notion of $T$-partial strategy. For a strategy $s_{i} \in S_{i}$ and a tree $T \in \mathbf{T}$, we denote by $s_{i}^{T}$ the strategy in the $T$-partial game induced by $s_{i}$ (i.e., $s_{i}^{T}\left( h_{i}\right) = s_{i}\left(h_{i}\right)$ for every information set $h_{i} \in H_i^T$ of player $i$ in the $T$-partial game). Denote by $S_i^T$ player $i$'s set of $T$-partial strategies.

A \emph{mixed strategy} of player $i \in I^0$, $\sigma_{i} \in \Delta(S_{i})$, in game $\Gamma$ specifies a probability distribution over player $i$'s set of pure strategies. With this notation, we let $\sigma_0$ the probability distribution over ``strategies'' of nature. We do not consider mixed strategies as an object of choice of players; this notion is just used here in technical ways.

A \emph{behavior strategy} for player $i \in I$ in game $\Gamma$,
\begin{equation*} \pi_i \in \Pi_i := \prod_{h_i \in H_i} \Delta(A_i(h_i))
\end{equation*}
is a collection of independent probability distributions, one for each of player $i$'s information set $h_{i}\in H_{i}$, where $\pi_i(h_i)$ specifies a mixed action in $\Delta(A_{h_i})$. With the behavior strategy $\pi_i$, define player $i$'s mixed action at node $n \in D_i^{T_n}$ to be $\pi_i(h_i(n))$. Thus, the behavior strategy $\pi_i$ specifies the mixed action of player $i \in I$ at each of her active decision nodes $n \in D_i^{T_n}$, both in the case that $n \in h_i(n)$ and in the case that $h_i(n)$ is a subset of nodes of a tree less expressive than $T_n$. It may be the case that $A_i(n) \supsetneqq A_i(h_i(n))$. Yet, we have automatically that $\pi_i$ does not assign strict positive probabilities to actions in $A_n \setminus A_{h_i(n)}$. (I.e., at the decision node $n$ of the richer tree $T_n$ player $i$ may have more actions than she is aware of at $h_i(n)$. In such a case, she is unable to use actions that she is unaware of.) With respect to nature, we let $\pi_0 \in \Pi_0 = \prod_{n \in D_0} \Delta(A_0(n))$.

For a behavior strategy $\pi_{i} \in \Pi_{i}$ and a tree $T \in \mathbf{T}$, we denote by $\pi_{i}^{T}$ the strategy in the $T$-partial game induced by $\pi_{i}$ (i.e., $\pi_{i}^{T}\left( h_{i}\right) = \pi_{i}\left(h_{i}\right)$ for every information set $h_{i} \in H_i^T$ of player $i$ in the $T$-partial game). Denote by $\Pi_i^T$ player $i$'s set of $T$-partial behavior strategies.

\subsection{Information Sets Consistent with Strategies}

In extensive-form games with unawareness there are two distinct notions of a strategy profile being consistent with a node that we call a ``strategy reaching a node'' and ``a node occurs with a strategy'', respectively. The first is a more ``subjective'' notion capturing what nodes a player with a certain awareness level expects a strategy profile to reach. The second notion is more an ``objective'' notion of what nodes actually occur with a strategy profile. Both notions are relevant. The first is relevant to extensive-form rationalizability, the second for self-confirming equilibrium.

We say that a strategy profile $s=\left( s_{j}\right) _{j\in I^0}\in S$ \emph{reaches a node} $n\in T$ if the players' actions and nature's moves $\left(s_{j}^{T}\left( h_{j}\left( n^{\prime }\right) \right)\right)_{j \in P(n')}$ in nodes $n^{\prime }\in T$ lead to $n$. That is, the sequence of action profiles induced by $s$ at predecessors of $n$ in $T$ lead to $n \in T$. Notice that by property (I4) (``no imaginary actions''), $s_{j}^{T}\left( h _{j}\left( n^{\prime }\right) \right) _{j\in I}$ is indeed well defined: even if $h_{j}\left(n'\right) \nsubseteq T$ for some $n^{\prime }\in T$, $\left(s_{j}^{T}\left( h_{j}\left( n'\right) \right)\right)_{j \in P(n')}$ is a profile of actions which is actually available in $T$ to the active players $j \in P(n')$ and possibly nature at $n'$. We say that a strategy profile $s\in S$ \emph{reaches} the information set $h_{i}\in H_{i}$ if $s$ reaches some node $n\in h_{i}$. The notion of reach is extended to individual strategies $s_{i}$ and opponent's strategy profiles $s_{-i}$ in an obvious way (see Schipper, 2018). For each player $i \in I$, denote by $H_i(s)$ the set of information sets of $i$ that are reached by the strategy profile $s$. This set may contain information sets in more than one tree.

We say that node $n \in \bar{T}$ in the upmost tree $\bar{T}$ \emph{occurs} with strategy profile $s=\left( s_{j}\right)_{j\in I^0}\in S$ if the players' actions and nature's moves $\left(s_{j}\left(h_{j}(n') \right)\right)_{j \in P(n')}$ in nodes $n' \in \bar{T}$ reach $n \in \bar{T}$. We extend the notion to any node in any tree by saying that node $n \in T$ occurs with strategy profile $s = \left(s_{j}\right)_{j\in I}\in S$ if $n' \in \bar{T}$ with $n'_T = n$ occurs with $s$. This is well-defined because $\mathbf{T}$ is a join semi-lattice. In particular, for any $T \in \mathbf{T}$ and $n \in T$ there is a node $n' \in \bar{T}$ such that $n'_T = n$. We say that information set $h_{i}\in H_{i}$ occurs with strategy profile $s \in S$ if some node $n \in \mathbf{D}_i$ with $h_i(n) = h_i$ occurs with $s$. Note that for this definition we do not require $n \in h_i$. The notion of occurs is extended to individual strategies $s_{i}$ and opponent's strategy profiles $s_{-i}$ in an obvious way (see Schipper, 2018). For each player $i \in I$, denote by $\tilde{H}_i(s)$ the set of information sets of $i$ that occur with strategy profile $s$. This set may contain information sets in more than one tree.

The notions of reaching nodes/information sets and nodes/information sets occurring are discussed further in Schipper (2018) who also provides examples as to where they differ.

We extend the definitions of information set reached and information sets occurring to behavior strategies in the obvious way by considering nodes/information sets that are reached/occurring with strict positive probability. For any $i \in I$, we let $H_i(\pi)$ denote the set of player $i$'s information sets that are reached with strict positive probability by the behavior strategy profile $\pi$ and $\tilde{H}_i(\pi)$ denote the set of player $i$'s information sets that occur with strict positive probability with the behavior strategy profile $\pi$.

\subsection{Belief Systems}

A \emph{belief system} of player $i \in I$ in game $\Gamma$,
\begin{equation*}
\beta_{i}=\left( \beta_{i}\left( h_{i}\right) \right) _{h_{i}\in H_{i}} \in \prod_{h_{i}\in H_{i}}\Delta \left( S_{-i}^{T_{h_{i}}}\right)
\end{equation*}
is a profile of beliefs -- a belief $\beta_{i}\left( h_{i}\right) \in \Delta \left( S_{-i}^{T_{h_{i}}}\right)$ about the other players' strategies (and possibly nature) in the $T_{h_{i}}$-partial game, for each information set $h_{i}\in H_{i}$, with the following properties:

\begin{itemize}
\item $\beta_{i}\left( h_{i}\right)$ reaches $h_{i}$, i.e., $\beta_{i}\left( h_{i}\right) $ assigns probability 1 to the set of strategy profiles of the other players (including possibly nature) that reach $h_{i}$.

\item If $h_{i}$ precedes $h_{i}^{\prime }$ (i.e., $h_{i}\rightsquigarrow h_{i}^{\prime }$), then $\beta_{i}\left( h_{i}^{\prime} \right)$ is derived from $\beta_{i}\left( h_{i}\right)$ by conditioning whenever possible.
\end{itemize}

Note that different from Heifetz, Meier, and Schipper (2013) a belief system specifies also beliefs about strategies of opponents and nature at information sets of terminal nodes. This is an essentially feature that we require for defining self-confirming equilibrium. Denote by $B_{i}$ the set of player $i$'s belief systems.\footnote{In some applications, we may want to fix prior beliefs about moves of nature. In such a case, we would consider $B_i$ to consist only of belief systems in which for every belief the marginal on nature is consistent with the fixed prior belief about moves of nature. Note that since a poorer tree may lack some moves of nature of a richer tree, further conditions on the prior may be imposed so as to form a system of priors, one for each tree, that is consistent across trees.}

For a belief system $\beta_{i}\in B_{i}$, a strategy $s_{i}\in S_{i}$ and an information set $h_{i}\in H_{i},$ define player $i$'s expected payoff \emph{at} $h_{i}$ to be the expected payoff for player $i$ in $T_{h_{i}}$ given $\beta_{i}\left( h_{i}\right) $, the actions prescribed by $s_{i}$ at $h_{i}$ and its successors, assuming that $h_{i}$ has been reached.

We say that with the belief system $\beta_{i}$ and the strategy $s_{i}$ player $i$ is \emph{rational} at the information set $h_{i}\in H_{i}^{\mathbf{D}}$ if either $s_{i}$ does not reach $h_{i}$ or there exists no strategy $s_{i}^{\prime }$ which is distinct from $s_{i}$ only at $h_{i}$ and/or at some of $h_{i}$'s successors in $T_{h_{i}}$ and yields player $i$ a higher expected payoff in the $T_{h_{i}}$-partial game given the belief $\beta_{i}\left( h_{i}\right)$ on the other players' strategies $S_{-i}^{T_{h_{i}}}$.

Player $i$'s \emph{belief system on behavior strategies} of opponents in game $\Gamma$,
\begin{eqnarray*}
\mu_i = \left(\mu_i(h_i)\right)_{h_i \in H_i} \in \prod_{h_i \in H_i} \Delta(\Pi_{-i}^{T_{h_i}})
\end{eqnarray*} is a profile of beliefs -- a belief $\mu_i(h_i) \in \Delta(\Pi_{-i}^{T_{h_i}})$ about the behavior strategies of other players (incl. possibly nature) in the $T_{h_i}$-partial game, for each information set $h_i \in H_i$, with the following properties
\begin{itemize}
\item $\mu_{i}\left( h_{i}\right)$ reaches $h_{i}$, i.e., $\mu_{i}\left( h_{i}\right) $ assigns probability 1 to the set of behavior strategy profiles of the other players (incl. possibly nature) that reach $h_{i}$.

\item If $h_{i}$ precedes $h_{i}^{\prime }$ (i.e., $h_{i}\rightsquigarrow h_{i}^{\prime }$), then $\mu_{i}\left( h_{i}^{\prime} \right)$ is derived from $\mu_{i}\left( h_{i}\right)$ by conditioning whenever possible.
\end{itemize} We denote by $M_i$ the set of player $i$'s belief systems over behavior strategies of opponents.

For a belief system $\mu_{i}\in M_{i}$, a behavior strategy $\pi_{i} \in \Pi_{i}$ and an information set $h_{i}\in H_{i}$, define player $i$'s expected payoff \emph{at} $h_{i}$ to be the expected payoff for player $i$ in $T_{h_{i}}$ given $\mu_{i}\left( h_{i}\right)$, the mixed actions prescribed by $\pi_{i}$ at $h_{i}$ and its successors, assuming that $h_{i}$ has been reached.

We say that with the belief system $\mu_{i}$ and the behavior strategy $\pi_{i}$ player $i$ is \emph{rational} at the information set $h_{i}\in H_{i}^{\mathbf{D}}$ if either $\pi_{i}$ does not reach $h_{i}$ or there exists no behavior strategy $\pi_{i}'$ which is distinct from $\pi_{i}$ only at $h_{i}$ and/or at some of $h_{i}$'s successors in $T_{h_{i}}$ and yields player $i$ a higher expected payoff in the $T_{h_{i}}$-partial game given the belief $\mu_{i}\left( h_{i}\right)$ on the other players' behavior strategies $\Pi_{-i}^{T_{h_{i}}}$.

\subsection{Self-Confirming Equilibrium\label{sce_section}}

The discussion of the first example in Section~\ref{Example} in the introduction made clear that the challenge for a notion of equilibrium is to deal with changes of awareness along the equilibrium paths. In a ``steady-state of conceptions'', awareness should not change. We incorporate this requirement into our definition of self-confirming equilibrium. For simplicity, we first consider a notion of self-confirming equilibrium in pure strategies.

\begin{defin}[Self-confirming equilibrium in pure strategies]\label{sce} A strategy profile $s \in S$ is a self-confirming equilibrium of $\Gamma$ if for every player $i \in I$:
\begin{itemize}
\item[(0)] Awareness is self-confirming along the path: There is a tree $T \in \mathbf{T}$ such that for all occurring information sets $h_i \in \tilde{H}_i(s)$ we have $h_i \subseteq T$.
\end{itemize}
\noindent There exists a belief system $\beta_i \in B_i$ such that
\begin{itemize}
\item[(i)] Players are rational along the path: With belief system $\beta_i$, strategy $s_i$ is rational at all occurring information sets in $\tilde{H}_i(s)$.
\item[(ii)] Beliefs are self-confirming along the path: For any information set of terminal nodes $h_i \in H_i^{\mathbf{Z}} \cap \tilde{H}_i(s)$ occurring with strategy profile $s$, the belief system $\beta_i$ is such that $\beta_i(h_i)$ assigns probability 1 to the subset of profiles of opponents' and nature's strategies of $S_{-i}^{T_{h_i}}$ that reach $h_i$. Moreover, for any preceding (hence non-terminal) information set $h_i' \rightsquigarrow h_i$, $\beta_i(h_i') = \beta_i(h_i)$.
\end{itemize}
\end{defin}

Condition (0) requires that awareness is constant along the equilibrium path. Players do not discover anything novel in equilibrium play. This is justified by the idea of equilibrium as a stationary rest-point or stable convention of play. Implicitly, it is assumed that discoveries if any are made before equilibrium is reached.

Condition (i) is a basic rationality requirement of equilibrium. Note that rationality is required only along information sets that occur along the path of play induced by the equilibrium strategy profile. The equilibrium notion is silent on off-equilibrium information sets (in particular on information sets that could be visited with $s_i$ but are not visited with $s_{-i}$). Condition (i) does not require that players believe others are rational along the path, believe that others believe that etc. It is just a ``minimal'' rationality requirement in an extensive-form game.

Condition (ii) consists of two properties. First, at the end of the game the player is certain of strategies of opponents and nature that allow her to reach the particular end of the game. That is, terminal beliefs are consistent with what has been observed during play (and hence at the end of the play). Second, beliefs do not change during the play. That is, beliefs at any information set reached during the play are consistent with what is observed at any point during the play and in particular with what is observed at the end of the game. Again, the idea is that everything that could have been learned on this path has been learned already in the past. This is justified by the idea of equilibrium as a stationary rest-point or stable convention of play as a result of prior learning. Note that this notion of equilibrium is silent on beliefs off equilibrium path.

It should be obvious that pure self-confirming equilibria may not exist even in standard games. Consider as a simple counterexample the matching pennies game, which fits our framework as we allow for simultaneous moves. As a remedy, we consider also an analogous notion of self-confirming equilibrium in behavior strategies.

\begin{defin}[Self-confirming equilibrium in behavior strategies]\label{sceb} A behavior strategy profile $\pi \in \Pi$ is a self-confirming equilibrium of $\Gamma$ if for every player $i \in I$:
\begin{itemize}
\item[(0)] Awareness is self-confirming along the path: There is a tree $T \in \mathbf{T}$ such that for all of player $i$'s visited information sets $h_i \in \tilde{H}_i(\pi)$ we have $h_i \subseteq T$.
\end{itemize}
\noindent There exists a belief system $\mu_i \in M_i$ such that
\begin{itemize}
\item[(i)] Players are rational along the path: With belief system $\mu_i$, behavior strategy $\pi_i$ is rational at all visited information sets in $\tilde{H}_i(\pi)$.
\item[(ii)] Beliefs are self-confirming along the path: For any information set of terminal nodes $h_i \in H_i^{\mathbf{Z}} \cap \tilde{H}_i(\pi)$ occurring with the behavior strategy profile $\pi$, the belief system $\mu_i$ is such that $\mu_i(h_i)$ assigns probability 1 to $\{\pi'_{-i} \in \Pi_{-i}^{T_{h_i}} : \pi'_j(h_j) = \pi_j(h_j)$ for $j \in I^0 \setminus \{i\}$ and $h_j \in \tilde{H}^{T_{h_i}}_j(\pi)\}$. Moreover, for any preceding (hence non-terminal) information set $h_i' \rightsquigarrow h_i$, $\mu_i(h_i') = \mu_i(h_i)$.
\end{itemize}
\end{defin}

The interpretation of properties (0) to (ii) is analogous to previous Definition~\ref{sce}. For property (ii), note that $\left\{\pi'_{-i} \in \Pi_{-i}^{T_{h_i}} : \pi'_j(h_j) = \pi_j(h_j) \mbox{ for } j \in I^0 \setminus \{i\} \mbox{ and } h_j \in \tilde{H}^{T_{h_i}}_j(\pi)\right\}$ is the set of behavior strategy profiles of opponents of player $i$ and nature that are behaviorally indistinguishable from $\pi$ at all information sets conceived by $i$ and found relevant by $i$ along actual paths of play induced by $\pi$ in $i$'s model.

We do not require that in self-confirming equilibrium player $i$ believes that opponents mix independently. This is because we do not find independence easy to motivate. The literature on self-confirming equilibrium knows both assumptions. For instance, independence is assumed in Fudenberg and Levine (1993a) but not in Rubinstein and Wolinsky (1994).

It is well-known in the literature (see for instance Fudenberg and Levine, 1993, Fudenberg and Kreps, 1995, Battigalli and Guaitoli, 1997, Kamada, 2010, Shimoji, 2012) that self-confirming equilibrium is a coarsening of Nash equilibrium in standard games. Nevertheless, in finite games with unawareness they may not exist due to failure of condition (0). The game in Figure~\ref{example1a} in Section~\ref{Example} constitutes a  simple counterexample.\\

\noindent \textbf{Example 1 (continued): Failure of existence of self-confirming equilibrium in finite extensive-form games with unawareness.} Condition (i), rationality along the path, requires that player 1 chooses $\ell_1$ and player 2 chooses $m_2$ in $\bar{T}$ and $r_2$ in $T$ in the game of Figure~\ref{example1a}. It is also easy to see that for each player there exists a belief system satisfying condition (ii) of the definition of self-confirming equilibrium. Yet, the play emerging from rational strategies reaches an information set of player 1 that contains a terminal node in $\bar{T}$ after being initially only aware of $T$, which violates condition (0). That is, awareness is not self-confirming along the path. Hence, there is no self-confirming equilibrium. The failure is due changes of awareness on any potential equilibrium path violating condition (0). The point is that not every extensive-form game with unawareness allows for a rational path of play with constant awareness of players.\hfill $\Box$\\

Consider a special subclass of extensive-form games with unawareness: An extensive-form game with unawareness has \emph{common constant awareness} if there exists $T \in \mathbf{T}$ such that for all $n \in \bar{T}$, $h_i(n) \subseteq T$ for all $i \in I$. That is, the awareness among players is constant and common at every play of the game. It does not imply though that players are fully aware of every move because we may have $T \prec \bar{T}$. The proof of the following remark is contained in the appendix.

\begin{rem}\label{existence} Consider a finite extensive-form game with unawareness. If it has common constant awareness then it possesses a self-confirming equilibrium. The converse is false.
\end{rem}

Special subclasses of extensive-form games with common constant awareness are games with full awareness, i.e., for all $n \in \bar{T}$ we have $h_i(n) \subseteq \bar{T}$, as well as standard extensive-form games.

\section{Discovery Processes and Equilibrium\label{discovery_section}}

In this section, we like to model how players playing a game discover actions and may eventually reach a self-confirming game.

Let $\mathbf{\Gamma}$ be the set of all extensive-form games with unawareness for which the initial building block (i.e., outlining the physical moves) is the finite extensive-form game with perfect information $\langle I, \bar{T}, P, (u_i)_{i \in I}\rangle$. By definition, $\mathbf{\Gamma}$ is finite.

For any extensive-form game with unawareness $\Gamma \in \mathbf{\Gamma}$, denote by $S_{\Gamma}$ the set of pure strategy profiles in $\Gamma$.

\begin{defin}[Discovered version]\label{discovered_version} Given an extensive-form game with unawareness $\Gamma = \langle I, \mathbf{T}, P, (H_i)_{i \in I}, (u_i)_{i \in I} \rangle \in \mathbf{\Gamma}$ and a strategy profile in this game $s_{\Gamma}$, the discovered version $\Gamma' = \langle I', \mathbf{T}', P', (H_i')_{i \in I'}, (u'_i)_{i \in I'} \rangle$ is defined as follows:
\begin{itemize}
\item[(i)] $I' = I$, $\mathbf{T}' = \mathbf{T}$, $P' = P$, and $u'_i = u_i$ for all $i \in I'$.

\item[(ii)] For $i \in I'$, the information sets in $H_i'$ of $\Gamma'$ are defined as follows: Let $$T^i_{s_{\Gamma}} := \sup\left\{T \in \mathbf{T} : h_i(n) \subseteq T, h_i(n) \in \tilde{H}_i(s_{\Gamma})\right\}.\footnote{This least upper bound exist in $\mathbf{T}$ since $\mathbf{T}$ is a finite (hence join-complete) join-semilattice.}$$

    For any $n \in \bar{T}$ (i.e., the upmost tree in $\mathbf{T}$) with $h_i(n) \in H_i$, $h_i(n) \subseteq T'$, $T', T'' \in \mathbf{T}$,
    \begin{itemize}
    \item[a.] if $T' \preceq T^i_{s_{\Gamma}} \preceq T'' \preceq \bar{T}$, the information set $h_i'(n_{T''}) \in H_i'$ is defined by\footnote{As defined previously, we take $n_{T''}$ the copy of node $n \in \bar{T}$ in the tree $T''$. When $T'' \equiv \bar{T}$, then $n_{T''} = n$.}
        $$h_i'(n_{T''}) := \left\{n' \in T^i_{s_{\Gamma}}: h_i(n') = h_i(n)\right\}.$$

    \item[b.] if $T' \preceq T'' \preceq T^i_{s_{\Gamma}}$, the information set $h_i'(n_{T''}) \in H_i'$ is defined by $$h_i'(n_{T''}) := \left\{n' \in T'': h_i(n') = h_i(n)\right\}.$$

    \item[c.] Otherwise, if $T' \not\preceq T^i_{s_{\Gamma}}$ and $T' \preceq T'' \preceq \bar{T}$, the information set $h_i'(n_{T''}) \in H_i'$ is defined by $$h_i'(n_{T''}) := h_i(n_{T''}).$$

    \end{itemize}

\end{itemize}
\end{defin}

When an extensive-form game with unawareness $\Gamma$ is played according to a strategy profile $s_{\Gamma}$, then some players may discover something that they were previously unaware of. The discovered version $\Gamma'$ of an ``original'' extensive-form game with unawareness $\Gamma$ represents the views of the players after the extensive-form game with unawareness has been played according to a strategy profile $s_{\Gamma}$. The discovered version has the same set of players, the same join-semilattice of trees, the same player correspondence, and the same payoff functions as the original game (condition (i)). What may differ are the information sets. In particular, in a discovered version some players may from the beginning be aware of more actions than in the ``original'' game $\Gamma$ but only if in the ``original'' game it was possible to discover these action with the strategy profile $s_{\Gamma}$. The information sets in the discovered version reflect what players have become aware when playing $\Gamma$ according to $s_{\Gamma}$.

To understand part $(ii)$ of Definition~\ref{discovered_version}, note first that $T^i_{s_{\Gamma}}$ is the tree that represents the player $i$'s awareness of physical moves in the game $\Gamma$ after it has been played according to strategy profile $s_{\Gamma}$. It is determined by the information sets of player $i$ that occur along the play-path in the upmost tree according to $s_{\Gamma}$. Now consider all information sets of player $i$ that arise at nodes in the upmost tree $\bar{T}$ in the ``original'' game $\Gamma$. These information sets may be on less expressive trees than $T^i_{s_{\Gamma}}$. Since player $i$ is now aware of $T^i_{s_{\Gamma}}$, all those information sets that in $\Gamma$ were on a tree less expressive than $T^i_{s_{\Gamma}}$ are now lifted to tree $T^i_{s_{\Gamma}}$, the tree that in player $i$'s mind represents the physical moves of the strategic situation after $\Gamma$ has been played according to $s_{\Gamma}$. Yet, this holds not only for reached nodes in the upmost tree $\bar{T}$ but also for copies of those nodes in trees $T'' \in \mathbf{T}$, $T^i_{s_{\Gamma}} \preceq T'' \preceq \bar{T}$. This is because by Property U4 of extensive-form games with unawareness the information sets at copies of those nodes in trees $T''$ are also on trees less expressive than $T^i_{s_{\Gamma}}$ in $\Gamma$. When information sets are lifted to higher trees, they contain all nodes in such a tree that previously gave rise to the information set at a lower tree in $\Gamma$. This explains part $a.$ of $(ii)$ of Definition~\ref{discovered_version}.

Part $b.$ pertains to information sets in trees less expressive than $T^i_{s_{\Gamma}}$. These are trees that miss certain aspects of tree $T^i_{s_{\Gamma}}$. These trees are relevant to player $i$ in $T'$ nevertheless as she has to consider other players' views of the strategic situation, their views of her view etc. Since other players may be unaware of aspects of which player $i$ is aware, she should consider her own ``incarnations'' with less awareness. In the discovered version $\Gamma'$, the information sets on trees $T'' \preceq T^i_{s_{\Gamma}}$ model the same knowledge of events as in information sets on tree $T^i_{s_{\Gamma}}$ provided that she is still aware of those events in $T''$. This is crucial for the discovered version to satisfy property U5 of extensive-form games with unawareness.

Part $c.$ just says that in the discovered version information set of player $i$ in trees incomparable to $T^i_{s_{\Gamma}}$ remain identical to the original game $\Gamma$. These information sets represent awareness that necessarily has not been discovered when $\Gamma$ is played according to strategy profile $s_{\Gamma}$.
\begin{figure}[h!]
\caption{Original $\Gamma$ (left game form) and Discovered Version $\Gamma'$ (right game form)\label{discovered}}
\begin{center}
\includegraphics[scale=0.5]{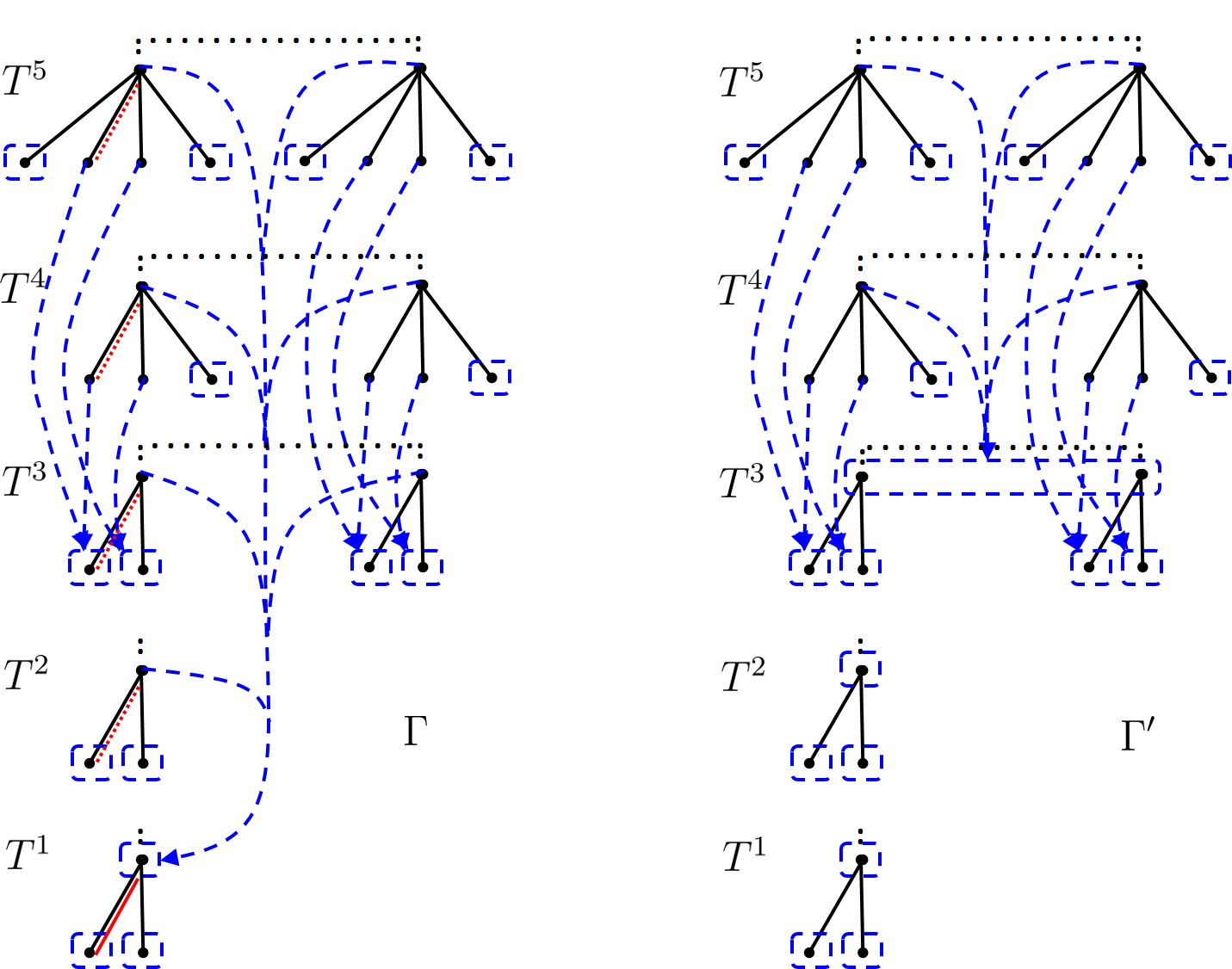}
\end{center}
\end{figure}

The notion of discovered version is illustrated in Figure~\ref{discovered}. This example is sufficiently rich to cover at least cases a. and b. distinguished in (ii) of Definition~\ref{discovered_version}. Consider the extensive-form game with unawareness to the left, $\Gamma$, as the ``original'' game. Initially, the player's awareness is given by tree $T^1$. The strategy in the original game is indicated by the solid orange line in $T^1$. It induces actions also in more expressive trees indicated by the orange dotted lines. When the strategy is executed, the player's awareness is raises to tree $T^3$. (I.e., the information set at the terminal node in $\bar{T} = T^5$ that occurs with the actions induced by the orange dotted strategy is located in $T^3$.)

The extensive-form game with unawareness to the right is the discovered version $\Gamma'$ if players (and possibly nature) follow the strategy profiles indicated by the orange dashed lines. Clearly, $T^5$ in Figure~\ref{discovered} corresponds to $\bar{T}$ in Definition~\ref{discovered_version} (ii), $T^1$ to $T'$, and $T^3$ to $T^i_{s_{\Gamma}}$. For case a., let $T''$ in Definition~\ref{discovered_version} (ii a.) be $T^4$. For case b., let $T''$ in Definition~\ref{discovered_version} (ii b.) correspond to $T^2$.

When a player discovers something, her awareness is raised. Consequently, discovered versions of a game involve more awareness. Game $\Gamma'$ has (weakly) more awareness than game $\Gamma$ if for every player and every node at which the player is active in $\Gamma'$, her information set is in a tree that is at least as expressive as the tree on which her corresponding information set is in $\Gamma$. The following definition makes this precise.

\begin{defin}[More awareness]\label{more_aware} Consider two extensive-form games with unawareness $\Gamma = \langle I, \mathbf{T}, P, (H_i)_{i \in I}, (u_i)_{i \in I} \rangle$ and $\Gamma' = \langle I', \mathbf{T}', P', (H_i')_{i \in I'}, (u'_i)_{i \in I'} \rangle$ with $I' = I$, $\mathbf{T}' = \mathbf{T}$, $P' = P$, and $u'_i = u_i$ for all $i \in I'$. $\Gamma'$ has (weakly) more awareness than $\Gamma$ if for every node $n$ and every active player $i \in P(n)$, $h_i(n) \subseteq T$ and $h'_i(n) \subseteq T'$ implies $T' \succeq T$.
\end{defin}

Discovered versions shall just reflect changes of awareness. The information about play, i.e., what players know about the history in the game, should not change. That is, in a discovered version players should have the same knowledge or ignorance about play modulo awareness as in the original game. Game $\Gamma'$ preserves information of game $\Gamma$ if $\Gamma'$ has weakly more awareness than $\Gamma$ and for each player and decision node of the player the information set contains the same nodes or copies thereof as the corresponding information set in $\Gamma$. The following definition makes this precise.

\begin{defin}[Information preservation]\label{information_preservation} Consider two extensive-form games with unawareness $\Gamma = \langle I, \mathbf{T}, P, (H_i)_{i \in I}, (u_i)_{i \in I} \rangle$ and $\Gamma' = \langle I', \mathbf{T}', P', (H_i')_{i \in I'}, (u'_i)_{i \in I'} \rangle$ with $I' = I$, $\mathbf{T}' = \mathbf{T}$, $P' = P$, and $u'_i = u_i$ for all $i \in I'$ such that $\Gamma'$ has (weakly) more awareness than $\Gamma$. $\Gamma'$ preserves information of $\Gamma$ if
\begin{itemize}
\item[(i)] for any $n$ and every active player $i \in P(n)$, $h_i(n)$ consists of copies of nodes in $h'_i(n)$.
\item[(ii)] for any tree $T \in \mathbf{T}$, any two nodes $n, n' \in T$ and every active player $i \in P(n) \cap P(n')$, if $h_i(n) = h_i(n')$ then $h'_i(n) = h'_i(n')$.
\end{itemize}
\end{defin}

\begin{prop}\label{well-defined} For any extensive-form game with unawareness $\Gamma \in \mathbf{\Gamma}$ and any strategy profile $s_{\Gamma} \in S_{\Gamma}$ in this game, the discovered version $\Gamma'$ is unique and an extensive-form game with unawareness. Moreover, $\Gamma'$ has more awareness than $\Gamma$. Finally, $\Gamma'$ preserves the information of $\Gamma$.
\end{prop}

The proof is relegated to the appendix. It verifies one-by-one the properties of extensive-form games with unawareness as well as Definitions~\ref{more_aware} and~\ref{information_preservation}.

Discovered versions do not depend on differences in strategies that are irrelevant to discoveries. They only depend on the realized path. This follows immediately from Definition~\ref{discovered_version} because $T^i_{s_\Gamma}$ depends only on $\tilde{H}_i(s_{\Gamma})$.

\begin{rem} For any extensive-form game with unawareness $\Gamma \in \mathbf{\Gamma}$, if $s_{\Gamma}$ and $s'_{\Gamma}$ are two strategy profiles in $\Gamma$ that generate the same path of nodes occurring in the upmost tree of $\Gamma$, then the discovered version given $s_{\Gamma}$ is identical to the discovered version given $s'_{\Gamma}$.
\end{rem}

The players interaction may lead to discoveries, interaction in the discovered games may lead to further discoveries etc. To model the set of discovery processes based on an extensive-form game with unawareness, we essentially define a stochastic game in which each state represents an extensive-form game with unawareness.

\begin{defin}[Discovery supergame] The discovery supergame based on $\mathbf{\Gamma}$ is the stochastic game $\langle \mathbf{\Gamma}, \tau \rangle$ defined as follows
\begin{itemize}

\item the set of states is a finite set of all extensive-form games with unawareness $\mathbf{\Gamma}$ (with identical initial building block $\langle I, \bar{T}, P, (u_i)_{i \in i} \rangle$.)

\item the transition probabilities are given by for $\Gamma, \Gamma' \in \mathbf{\Gamma}$, $s \in S_{\Gamma}$,
    $$\tau(\Gamma' \mid \Gamma, s) = \left\{\begin{array}{cl} 1 & \mbox{if } \Gamma' \mbox{ is the discovered version of } \Gamma \mbox{ given } s \\
    0 & \mbox{otherwise} \end{array}\right.$$

\end{itemize}
\end{defin}

The discovery supergame is a stochastic game in which states are identified with extensive-form games with unawareness all based on the same building blocks. This means that each stage-game is an extensive-form game with unawareness. The set of players is the set of players in the underlying extensive-form games with unawareness (including nature if any). Each player's set of actions is state-dependent and consist of the strategies at the extensive-form games with unawareness. Since $\Gamma$ and $\Gamma'$ differ in information sets, they also differ in sets of available strategies since strategies ascribe actions at information sets. The transition probabilities are degenerate in the sense that only transitions to discovered versions are allowed (given the strategy profiles) and by Proposition~\ref{well-defined} the discovered version is unique. Payoff functions of players are given by the underlying extensive-form games with unawareness. Since all those games have the same building block, payoff functions are in fact the same in all states. What changes from state to state are information sets (and hence also the set of strategies available at those stage-games as well as the belief systems).\footnote{This notion of stochastic game differs slightly from the common definition of stochastic game (cf. Shapley, 1953). First, in our setting there is an extensive-form game with unawareness at each state of the stochastic game while in standard stochastic games there is a matrix game at each state. Second, we do not assume that players maximize payoffs across stage-games. Third, in our setting the transition probabilities are degenerate.}

Clearly, the discovery supergame cannot be interpreted as a game that players are necessarily fully aware. Rather, it is a convenient model for the modeler/analyst. Consequently, the supergame strategies of player in this discovery supergame are not objects actually chosen by players but just conveniently summarize the modeler's assumptions about players' play in all those games. A \emph{discovery supergame strategy} of player $i$ in the discovery supergame $\langle \mathbf{\Gamma}, \tau \rangle$ is a mapping $f_i: \mathbf{\Gamma} \longrightarrow \bigcup_{\Gamma \in \mathbf{\Gamma}} \Delta\left(S_{\Gamma, i}\right)$ that assigns to each game $\Gamma \in \mathbf{\Gamma}$ a probability distribution over strategies of that player in this game $\Gamma$ (i.e., $f_i(\Gamma) \in \Delta(S_{\Gamma, i})$). The notion makes clear that we only consider discovery strategies that are stationary Markov strategies. For each player $i \in I^0$ (including nature) denote by $F_i$ the set of all discovery strategies and by $F = \times_{i \in I^0} F_i$. Denote by $f = (f_i)_{i \in I^0}$ a profile of discovery strategies. We extend the definition of transition probabilities in order to be able to write $\tau(\cdot \mid \Gamma, f)$ for any $\Gamma \in \mathbf{\Gamma}$ and $f \in F$. We allow $f_i$ to assign mixtures over strategies of player $i$ in order to capture the modeler's uncertainty over player $i$'s strategies.

\begin{defin}[Discovery process] A discovery process $\langle \mathbf{\Gamma}, \tau, (f_i)_{i \in I^0} \rangle$ consists of a discovery supergame $\langle \mathbf{\Gamma}, \tau \rangle$ and a discovery supergame strategy $f_i: \mathbf{\Gamma} \longrightarrow \bigcup_{\Gamma \in \mathbf{\Gamma}} \Delta\left(S_{\Gamma, i}\right)$, one for each player $i \in I^0$ (including nature if any).
\end{defin}

In our formulation, every discovery process is a Markov process. An extensive-form game with unawareness $\Gamma \in \mathbf{\Gamma}$ is an absorbing state of the discovery process $\langle \mathbf{\Gamma}, \tau, f \rangle$ if $\tau(\Gamma \mid \Gamma, f) = 1$.

\begin{defin}[Self-confirming game] An extensive-form game with unawareness $\Gamma \in \mathbf{\Gamma}$ is a self-confirming game of a discovery process $\langle \mathbf{\Gamma}, \tau, f \rangle$ if $\Gamma$ is an absorbing state of $\langle \mathbf{\Gamma}, \tau, f \rangle$.
\end{defin}

We believe that this terminology is justified by the fact in a self-confirming game play will not lead to further discoveries and changes of awareness and the information structure. All players' subjective representations of the game are in a steady-state. In this sense, the game is self-confirming.

It is easy to see that every discovery process leads to a self-confirming game. Suppose to the contrary there is a discovery process that is not absorbing, then there must exist a cycle since $\mathbf{\Gamma}$ is finite. In such a cycle there must exist two distinct extensive-form games with unawareness $\Gamma, \Gamma' \in \mathbf{\Gamma}$, $\Gamma \neq \Gamma'$, such that $\Gamma$ is discovered from $\Gamma'$ and $\Gamma'$ is discovered from $\Gamma$ (possibly via further games). Yet, by Proposition~\ref{well-defined}, discovered versions must have more awareness, awareness must weakly increase along any discovery process, a contradiction to a cycle. Note that ``weakly more awareness'' is a partial order on $\mathbf{\Gamma}$. If $\Gamma$ has weakly more awareness than $\Gamma'$ and $\Gamma'$ has weakly more awareness than $\Gamma$, then $\Gamma = \Gamma'$. This proves the following assertion:

\begin{prop}\label{absorbing} Every discovery process leads to a self-confirming game.
\end{prop}

\begin{rem} A discovery process may have more than one self-confirming game since discovery strategies allow for mixtures over stage-game strategies. E.g., consider Example 2 in the introduction and a discovery process in which the modeler considers possible that player 2 can take any strategy in the initial game given in Figure~\ref{example2a} but players adopt only rationalizable strategies in all other games. Then both the games in Figures~\ref{example2b} and~\ref{example2c} are self-confirming games of the discovery process.
\end{rem}

\subsection{Rationalizable Discovery Processes\label{rat_discovery_section}}

How to select among discovery processes? Which behaviorial assumptions should be imposed on discovery processes? Clearly, it would be absurd to assume that players chose optimal discovery strategies since this would presume awareness of the discovery supergame and hence awareness of everything modelled in $\mathbf{\Gamma}$. In other words, there would not be anything to discover.

We propose to restrict discovery processes to extensive-form rationalizable strategies in each $\Gamma \in \mathbf{\Gamma}$. A rational player in a novel game should be able to reason about the rationality of others, their (strong) beliefs about rationality etc. This selects among discovery processes and games that can be discovered in such processes. Figures~\ref{example2b} and~\ref{example2d} show two examples of self-confirming discovered versions of the game in Figure~\ref{example2a}. But there is an important difference. The self-confirming version in Figure~\ref{example2d} can never be discovered by playing rationally in the original game of Figure~\ref{example2a} or any discovered versions thereof. While the game of Figure~\ref{example2b} is the self-confirming game of the discovery process restricted to extensive-form rationalizable strategies in each game, the game of Figure~\ref{example2d} necessarily requires discovery processes to put strict positive probability to non-rationalizable strategies. These observations give rise to notions of rationalizable discovery processes and rationalizable discovered versions, i.e., discovery process and versions of games that can be discovered when players play extensive-form rationalizable strategies in the original game, any discovered game, any discovered game of a discovered game etc. Rationalizable versions of games emerge in discovery processes in which players play only extensive-form rationalizable strategies.

We use extensive-form rationalizability \`{a} la Pearce (1984) and Battigalli (1997) extended by Heifetz, Meier, and Schipper (2013) to extensive-form games with unawareness. It is an iterative reduction procedure of beliefs that allows for correlated beliefs about opponents' strategies. It captures common strong belief in rationality (Battigalli and Siniscalchi, 2002, Guarino, 2020).

\begin{defin}[Extensive-Form Rationalizable Strategies]\label{EFR} Fix an extensive-form game with unawareness. Define, inductively, the following sequence of belief systems and strategies of player $i \in I$:
\begin{eqnarray*}
B_{i}^{1} & = & B_{i} \\
R_{i}^{1} & = & \left\{ s_{i}\in S_{i}:
\begin{array}{l}
\mbox{there exists a belief system } \beta_{i} \in B_{i}^{1} \mbox{ with which
for every } \\
\mbox{information set } h_{i} \in H_{i} \mbox{ player } i \mbox{ is rational at } h_{i}
\end{array}
\right\} \\
& \vdots & \\
B_{i}^{k} & = & \left\{\beta_{i} \in B_{i}^{k-1}:
\begin{array}{l}
\mbox{for every information set } h_{i}, \mbox{ if there exists some } \\
\mbox{profile of the other players' strategies } \\ s_{-i} \in R_{-i}^{k-1} = \prod_{j
\neq i} R_{j}^{k-1} \mbox{ such }
\mbox{ that } s_{-i} \mbox{ reaches } h_{i}, \\ \mbox{ then } \beta_{i}(h_{i})
\mbox{ assigns probability } 1 \mbox{ to } R_{-i}^{k-1, T_{h_{i}}}
\end{array}
\right\} \\
R_{i}^{k} & = & \left\{ s_{i} \in S_{i}:
\begin{array}{l}
\mbox{ there exists a belief system } \beta_{i}\in B_{i}^{k} \mbox{ with which
for every } \\
\mbox{ information set } h_{i} \in H_{i} \mbox{ player } i \mbox{ is rational at } h_{i}
\end{array}
\right\}
\end{eqnarray*}
The set of player $i$'s extensive-form rationalizable strategies is
\begin{equation*}
R_{i}^{\infty } = \bigcap_{k=1}^{\infty} R_{i}^{k}.
\end{equation*}
\end{defin}

Letting nature's ``extensive-form rationalizable strategies'' be $R_0^{\infty} = S_0$, the set of extensive-form rationalizable strategy profiles is
\begin{equation*}
R^{\infty } = \times_{i \in I^0} R_i^{\infty}.
\end{equation*}

Denote by $R^{\infty}_{\Gamma, i}$ the set of extensive-form rationalizable strategies of player $i$ in the extensive-form game with unawareness $\Gamma$. A rationalizable discovery process is now defined as a discovery process where for each extensive-form game with unawareness each player is restricted to play only extensive-form rationalizable strategies in the game.

\begin{defin}[Rationalizable discovery process] A discovery process $\langle \mathbf{\Gamma}, \tau, (f_i)_{i \in I^0} \rangle $ is a rationalizable discover process if for all players $i \in I$, $f_i: \mathbf{\Gamma} \longrightarrow \bigcup_{\Gamma \in \mathbf{\Gamma}} \Delta\left(R^{\infty}_{\Gamma, i}\right)$.
\end{defin}

Again, we emphasize that player $i$'s discovery supergame strategy $f_i$ shall be interpreted as representing the analyst's belief in the strategies played by player $i$ in extensive-form games with unawareness in $\mathbf{\Gamma}$. To the extent to which $f_i(\Gamma)$ is a nondegenerate probability distribution in $\Delta\left(R^{\infty}_{\Gamma, i}\right)$, it represents the analyst's uncertainty about which extensive-form rationalizable strategy player $i$ plays in $\Gamma$. We also emphasize again that in a rationalizable discovery process players play extensive-form rationalizable strategies \emph{within} each game of the discovery process. At each game along the discovery process, players maximize expected payoffs just in that game.

Often, the analyst wants to analyze a particular game with unawareness. Thus, it will be helpful to designate it as the initial state of the discovery supergame. We denote by $\langle \mathbf{\Gamma}, \tau, \Gamma^0 \rangle$ the discovery supergame with \emph{initial game} $\Gamma^0$.

A self-confirming game that is a result of a rationalizable discovery process we call a \emph{rationalizable self-confirming game}. It is easy to see that for every game with unawareness there exists a rationalizable self-confirming game. By Proposition~\ref{absorbing}, we know that every discovery process is absorbing and thus leads to a self-confirming version. As a corollary, also every rationalizable discovery process must be absorbing and leading to a self-confirming version. The existence of a rationalizable discovery process follows now from Proposition 1 in Heifetz, Meier, and Schipper (2013), who show that for every (finite) extensive-form game with unawareness the set of extensive-form rationalizable strategies is nonempty for every player. Thus, we have the following observation:

\begin{prop}\label{existence_rscv} For every extensive-form game with unawareness $\Gamma^0$ there exists a rationalizable discovery process $\langle \mathbf{\Gamma}, \tau, \Gamma^0, (f_i) \rangle$ that leads to a self-confirming game. We call such a self-confirming game a rationalizable self-confirming game.
\end{prop}

\subsection{Equilibrium\label{equilibrium}}

In this section, we return to the quest for an appropriate equilibrium for extensive-form games with unawareness. Previously we argued that often extensive-form games with unawareness do not possess equilibria that capture the result of a learning process because of the self-destroying nature of games with unawareness. Yet, since for every extensive-form game with unawareness there exists a discovery process that leads to a self-confirming version, the appropriate notion of equilibrium of a game with unawareness should naturally involve the equilibrium in the self-confirming version. This captures equilibrium both of conceptions and behavior.

We also argued for restricting discovery processes to rationalizable discovery processes. This motivates to restrict equilibria to extensive-form rationalizable strategies as well since it would be odd to assume that players play extensive-form rationalizable strategies all along the discovery process but once a rationalizable self-confirming version is reached and a convention of play emerges, such equilibrium convention suddenly involves strategies that are not extensive-form rationalizable. That is, we propose to use extensive-form rationalizability not only to put endogenously restrictions on the games that can be discovered but also on the self-confirming equilibrium that may emerge in absorbing states of discovery processes. While self-confirming equilibrium is a rather weak solution concept, the requirement of using only extensive-form rationalizable strategies strengthens it considerably as extensive-form rationalizability involves forward induction.

A technical obstacle in defining self-confirming equilibrium in extensive-form rationalizable strategies is that self-confirming equilibrium is defined in behavior strategies whereas extensive-form rationalizability is defined in (pure) strategies. We do require the extension of self-confirming equilibrium to behavior strategies because it may not exist in pure strategies even in standard games. Although mixed strategies do not make sense in games with unawareness when interpreted as an object of choice of players because it would mean that players can mix ex ante over strategies involving actions that they might be unaware of ex ante, we can consider mixed strategies that are equivalent to behavior strategies from a modeler's view. This requires us to spell out the notion of equivalence between strategies.

For any node $n$, any player $i \in I^0$, and any opponents' profile of strategies $s_{-i}$ (including nature if any), let $\rho(n \mid \beta_i, s_{-i})$ and $\rho(n \mid \sigma_i, s_{-i})$ denote the probability that $(\beta_i, s_{-i})$ and $(\sigma_i, s_{-i})$ reach node $n$, respectively. For any player $i \in I^0$, a mixed strategy $\sigma_i$ and a behavior strategy $\beta_i$ are \emph{equivalent} if for every profile of opponents' strategies $s_{-i} \in S_{-i}$ and every node $n \in \mathbf{N}$ of the extensive-form game with unawareness $\rho(n \mid \sigma_i, s_{-i}) = \rho(n \mid \beta_i, s_{-i})$. This notion of equivalence between strategies is based on the notion of strategies reaching nodes. Schipper (2018) shows that it implies also equivalence between strategies with respect to nodes occurring.

\begin{defin}[Self-confirming equilibrium in extensive-form rationalizable conjectures]\label{rsce} A behavior strategy profile $\pi^* = (\pi_i^*)_{i \in I^0} \in \Pi$ is a self-confirming equilibrium in extensive-form rationalizable conjectures of the extensive-form game with unawareness $\Gamma$ if it is a self-confirming equilibrium of $\Gamma$ and for every player $i \in I^0$ any mixed strategy $\sigma^*_i$ equivalent to $\pi^*_i$ assigns zero probability to any strategy of player $i$ that is not extensive-form rationalizable.\footnote{We call our equilibrium notion self-confirming equilibrium in extensive-form rationalizable conjectures in order to distinguish it from different versions of rationalizable self-confirming equilibrium in Rubinstein and Wolinsky (1994), Esponda (2013), Dekel, Fudenberg, and Levine (1999), and Fudenberg and Kamada (2015, 2018).}
\end{defin}

It is easy to produce examples showing that self-confirming equilibrium in extensive-form rationalizable conjectues refines self-confirming equilibrium. The criterium of extensive-form rationalizability strongly refines self-confirming equilibrium because it imposes forward-induction embodied in extensive-form rationalizability. As an example, we discuss in Section~\ref{repeated_games} the self-confirming equilibrium in extensive-form rationalizable conjectures of the battle-of-the-sexes game with an outside option.

The following theorem asserts that for every finite extensive-form game with unawareness there exists a steady-state of conceptions and behavior emerging from rationalizable play.

\begin{theo}\label{main_theorem} For every extensive-form game with unawareness there exists a rationalizable discovery process leading to a rationalizable self-confirming game which possesses a self-confirming equilibrium in extensive-form rationalizable conjectures.
\end{theo}

The proof is contained in the appendix.

We interpret this result as a general existence result for ``stage-game'' equilibrium in \emph{conceptions and behavior} for finite extensive-form games with unawareness.

\begin{rem} Consider a (rationalizable) discovery process leading to a (rationalizable) self-confirming game with a self-confirming equilibrium (in extensive-form rationalizable conjectures). Then this is a self-confirming game with a self-confirming equilibrium of \emph{every} extensive-form game visited in the discovery process.
\end{rem}

\section{Discussion\label{discussion}}

\subsection{Payoff Uncertainty\label{payoff uncertainty}}

In games with unawareness, it is not necessarily the case that the game is common knowledge among players. Thus, also preferences are not necessarily common knowledge as some players may not even conceive of all terminal nodes. However, upon becoming aware, players are assumed to understand the \emph{possible} payoff consequences of the newly discovered terminal nodes. At a player's information set, the player may still be uncertain about the ultimate payoff consequences of herself or other players because histories with different payoff consequences may be contained in that information set. Yet, Property I7 requires that at the end of each stage-game the player observes her \emph{own} payoff but not necessarily the payoffs of other players. Since this is required in every game, this property is implicitly commonly known among all players. So each player may learn during the play of each stage-game about her own payoff and the payoffs of others but while she is certain of her own payoff at the end of the stage-game, she may still face uncertainty about opponents' payoffs. Learning about opponents' payoffs is facilitated by the best-rationalizability principle (Battigalli, 1996) embodied in extensive-form rationalizability applied to each game along the discovery process. By attributing to the greatest possible extent rationality and higher-order (strong) belief in rationality to opponents, a player is able to deduce information about opponents' payoffs.

One may argue that in some settings it is implausible to assume that a player goes from not conceiving of a terminal node and hence its associated payoffs to perfectly knowing her payoff at the end of the stage-game. For instance, when R\"{o}ntgen discovered X-rays, he did not realize immediately the negative side-effects of radiation due to X-rays. But if we were to allow payoff uncertainty over own payoffs at the end of our stage-game, then our stage-game would not model the entire relevant situation. This would be at odds with our focus on stage-game equilibrium. Thus, from a methodological point of view, we believe property I7 is justified.

In applications for game theoretic experiments, it may be useful to require the even stronger assumption of perfect observability of \emph{every} players' payoff at the end of the game or even perfect observability of terminal nodes but then allow for payoff uncertainty in a more ad hoc way via the solution concept. In particular, instead of requiring that players play extensive-form rationalizable strategies in every stage-game along the discovery process, one may require only that every player plays rationally.\footnote{We thank an anonymous reviewer for this suggestion.} Thus, no player would necessarily assume opponents' to play rational or believe that they believe that opponents' play rational etc. and hence would not make use of opponents' payoff information. Compared to requiring extensive-form rationalizability along the discovery process, this would enlarge the set of self-confirming games that eventually could be discovered. Yet, as Example 2 demonstrates it would still restrict the set of self-confirming games that could eventually been discovered.

\subsection{Finitely Repeated Games with Unawareness\label{repeated_games}}

Our framework does not allow for \emph{infinitely} repeated games as stage-games. Yet, finitely repeated games can be handled w.l.o.g. in our framework because we allow stage-games to be finite extensive-games with unawareness. Any finitely repeated game - no matter whether the stage-game is a normal-form game or an extensive-form game - is itself a finite extensive-form game. Moreover, our framework explicitly allows for simultaneous moves.

Below we demonstrate this with an example of a twice-repeated Battle-of-the-Sexes game with an outside option.\footnote{I thank Byung Soo Lee for suggesting such an example.}  This example also illustrates the role of forward-induction in games with unawareness and the refining power of self-confirming equilibrium in extensive-form rationalizable strategies.
\begin{figure}[h!]
\caption{Twice-Repeated Battle-of-the-Sexes Game with an Outside Option\label{BoS1}}
\begin{center}
\includegraphics[scale=0.33]{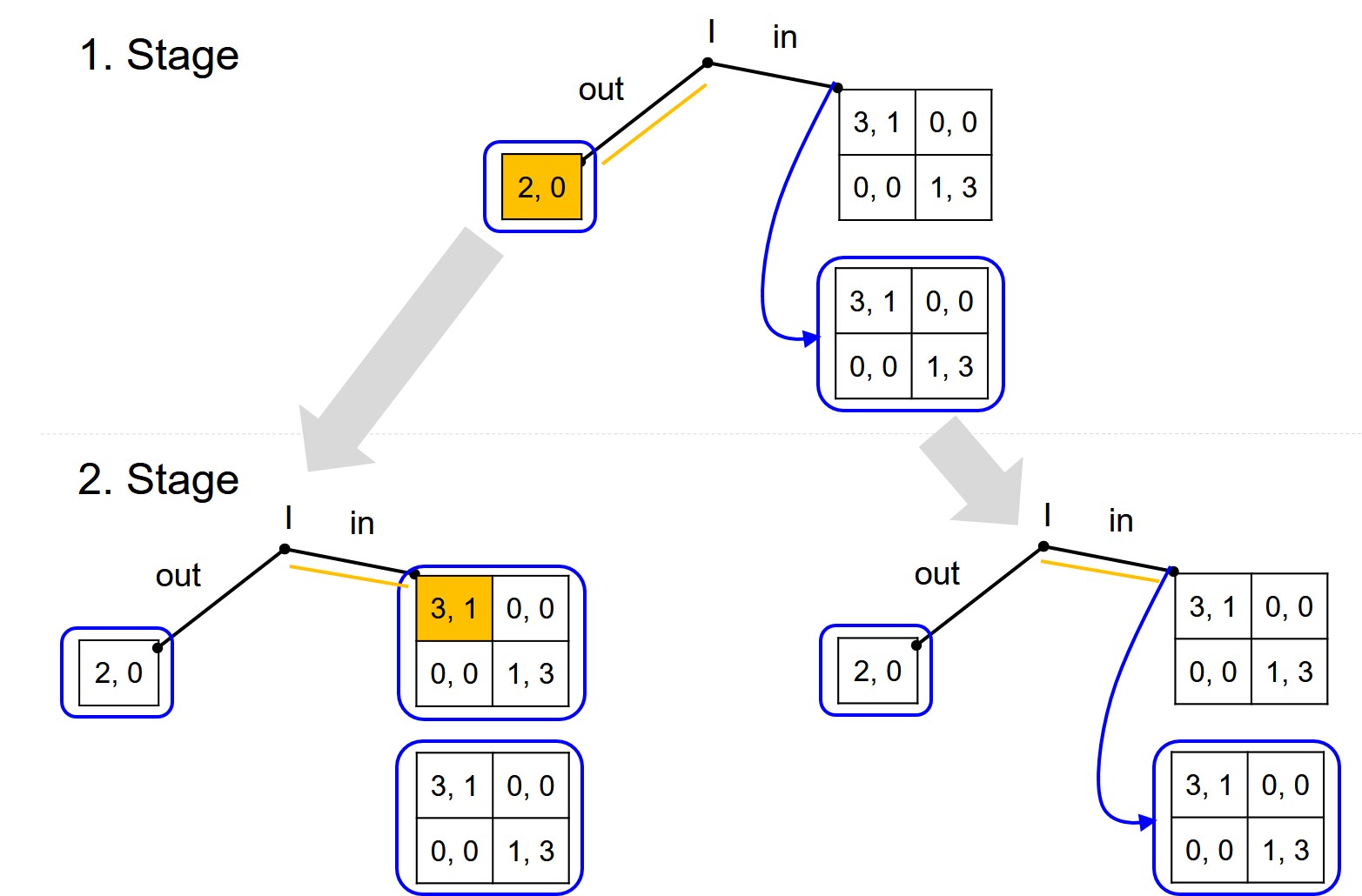}
\end{center}
\end{figure}

Initially player 2 is unaware of the outside option of player 1. Player 1 can make him aware of the outside option by taking it in the first stage and subsequently not taking it in the repetition. In such a case the presence of the outside option unfolds its power of forward-induction via extensive-form rationalizability embodied in our solution concept.

The game is described in Figure~\ref{BoS1}. We assume that throughout the interaction, player 1 is aware of everything. For simplicity we omit his information sets. The blue information sets model the information and awareness of player 2. There are two stages. The first stage features a game with two trees. The upper tree consists of a standard Battle-of-the-Sexes game with an outside option. If player 1 does not take the outside option, then players play a Battle-of-the-Sexes game. Player 2 is unaware of player 1's outside option. This is indicated by the blue information set that emanates after player 1 chooses ``in'' and that consists of a standard $2 \times 2$ Battle-of-the-Sexes game. This standard $2 \times 2$ game constitutes a tree less expressive than the upper tree. When player 1 chooses ``out'', then player 2 becomes aware of ``out'' as indicated by his blue information set at the terminal node with payoff vector $(2, 0)$.

The second stage of the repeated game depends on the action of player 1 in the first. When player 1 chooses ``out'' in the first stage, then player 2 becomes aware of the outside option. This is reflected by the updated information sets in the left game (stage 2). When player 1 chooses ``in'', then player 2 remains unaware of ``out''. Consequently, in the second stage the game is identical to the first stage (the right game of the stage 2 in Figure~\ref{BoS1}).

In this repeated game, it is extensive-form rationalizable for player 1 to play ``out'' in the first stage and (in, B) in the second. Moreover, player 2 plays B in the left game of the second stage. Initially, player 2 is unaware of ``out''. Thus, player 2 cannot forward-induce from player 1 not taking ``out'' that player 1 plays B in the Battle-of-the-Sexes. That is, unawareness of the outside option mutes forward-induction. Player 1 realizes this and takes ``out'' in the first stage so as to raise player 2's awareness of ``out''. In the second stage, he can now choose ``in'' anticipating that player 2 chooses B by forward induction.
\begin{figure}[h!]
\caption{Discovered Twice-Repeated Battle-of-the-Sexes Game with an Outside Option\label{BoS2}}
\begin{center}
\includegraphics[scale=0.33]{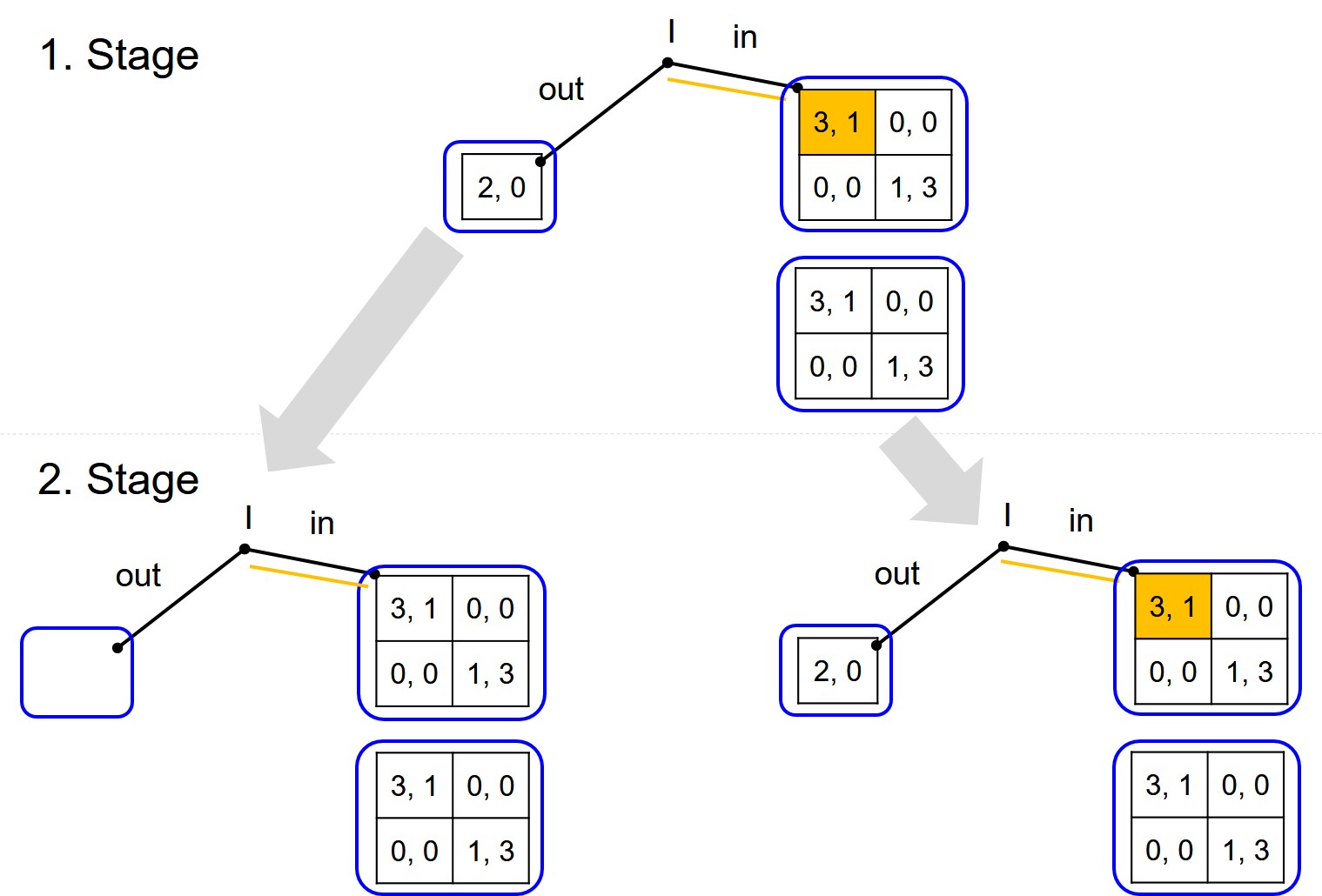}
\end{center}
\end{figure}

Once the twice-repeated game is over, the discovered version is depicted in Figure~\ref{BoS2}. It is now a repetition of the Battle-of-the-Sexes game with an outside option in which player 2 is aware of the player 1's outside option from the beginning. This repeated game is self-confirming. In this game, playing (in, B) in each stage for player 1 and playing B for player 2 is a self-confirming equilibrium in extensive-form rationalizable strategies.
\begin{figure}[h!]
\caption{Discovery Process of Twice-Repeated Battle-of-the-Sexes Game with an Outside Option\label{BoS_process}}
\begin{center}
\includegraphics[scale=0.7]{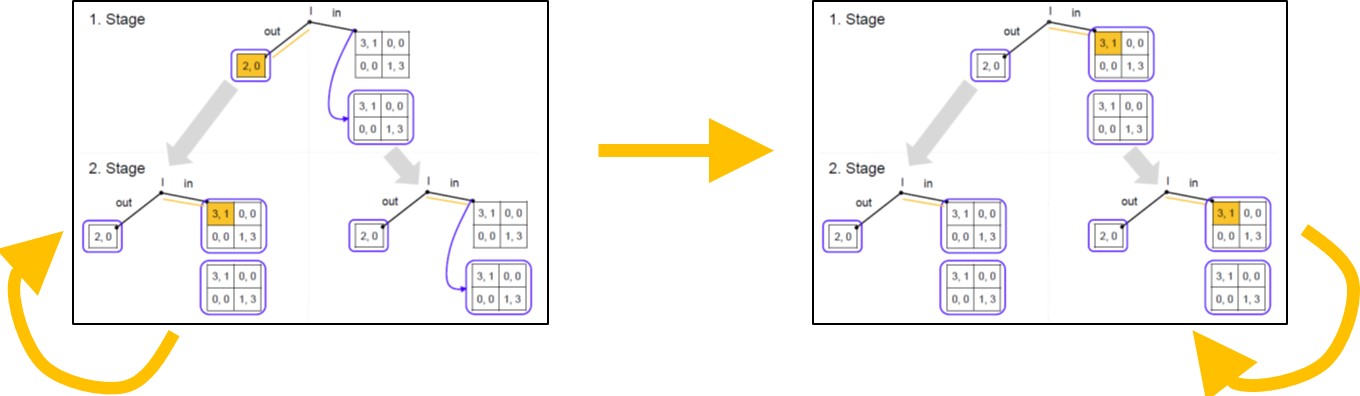}
\end{center}
\end{figure}

Note that the notion of discovered version is used in two ways in this example. Once in the left second stage game of the game in Figure~\ref{BoS1} and once in Figure~\ref{BoS2}. The latter is the discovered version of the \emph{repeated} game while the former is the discovered version of the stage game. The discovery process is depicted in Figure~\ref{BoS_process}.

The example demonstrates that the rationalizable discovery process features the first player initially taking the outside option so as to raise the awareness of player 2 and utilize the power of forward-induction in the latter stages by not taking the outside option. The self-confirming game is essentially the standard Battle-of-the-Sexes game with an outside option and the forward-induction outcome is unique self-confirming equilibrium in extensive-form rationalizable strategies.

\subsection{Mutual Belief of Constant Awareness}

\begin{figure}[h!]
\caption{No mutual belief of constant awareness along the equilibrium path\label{counterexample}}
\begin{center}
\includegraphics[scale=0.35]{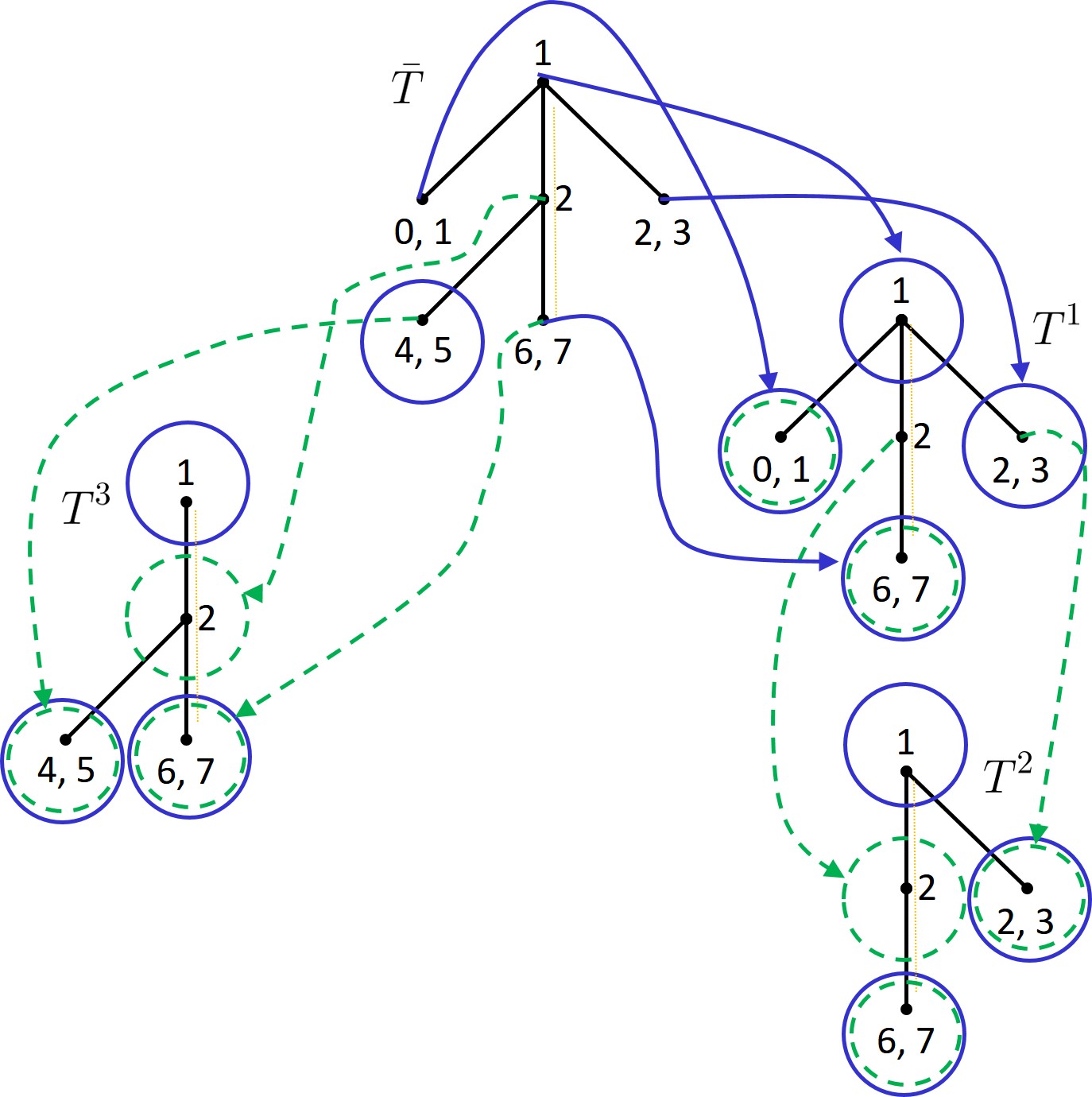}
\end{center}
\end{figure}

Our definition of self-confirming equilibrium requires among others that each player's awareness is constant along the equilibrium paths. One may ask whether this implies that there is mutual or even common belief in constant awareness. This would be a sensible requirement for equilibrium. Not only shall each player's awareness be constant in equilibrium but each player shall also believe that any other player's awareness is constant in equilibrium. Unfortunately this is not necessarily the case if players awareness is incomparable in equilibrium. Figure~\ref{counterexample} presents a counterexample. There are two players and four trees. The unique equilibrium path (which is also the unique extensive-form rationalizable path) is given by thin orange lines. The information structure of player 1 is drawn in blue, the one of player 2 is marked with dashed green lines. Player 1 is aware of both of his actions left and right but remains unaware of the left action of player 2. Player 2 is aware of his left action but remains unaware of player 1's left and right actions. Along the equilibrium path in the upmost tree $\bar{T}$ all of player 1's information sets are located in tree $T^1$. All of player 2's information sets along the equilibrium path in $\bar{T}$ are located in tree $T^3$. Thus both players have constant awareness. Yet, the information sets of player 2 along the equilibrium path in $T^1$ move from $T^2$ at player 2's decision node to $T^1$ at the terminal node. Player 1 believes that player 2's awareness along equilibrium is changing. Thus, there is an opportunity for further strengthening the equilibrium concept by requiring also mutual belief in constant awareness. For existence, we would need to restrict to extensive-form games with unawareness in which the set of trees form a complete lattice so that there is always a meet of all players' trees on which awareness could be constant for all players and in which this fact would be common belief.

\subsection{Awareness of Unawareness\label{awunaw}}

So far we focused on unawareness proper but did not discuss awareness of unawareness. More precisely, if a player is unaware of an action then she is unaware that she is unaware of it. Yet, a player may have a vague idea that she might be unaware of ``something'' (Schipper, 2021). An extension of the framework would be flexible enough to allow for awareness of unawareness of ``something''. As in previous work (Halpern and Rego, 2014, Heifetz, Meier, and Schipper, 2013), we can model awareness of unawareness by including imaginary actions as placeholders for actions that a player may be unaware and terminal nodes/evaluations of payoffs that reflect her awareness of unawareness. A player may now take such an imaginary action (or some other investigative action) that may or may not reveal what happens without being able to precisely anticipate what happens. We refer to Heifetz, Meier, and Schipper (2013, Section 2.6) for modeling details. Awareness of unawareness can be retained in self-confirming equilibrium of a self-confirming game only if in such an equilibrium the player decides not to investigate (further) her awareness of unawareness.
\begin{figure}[h!]
\caption{\label{awareness_unawareness0}}
\begin{center}
\includegraphics[scale=0.45]{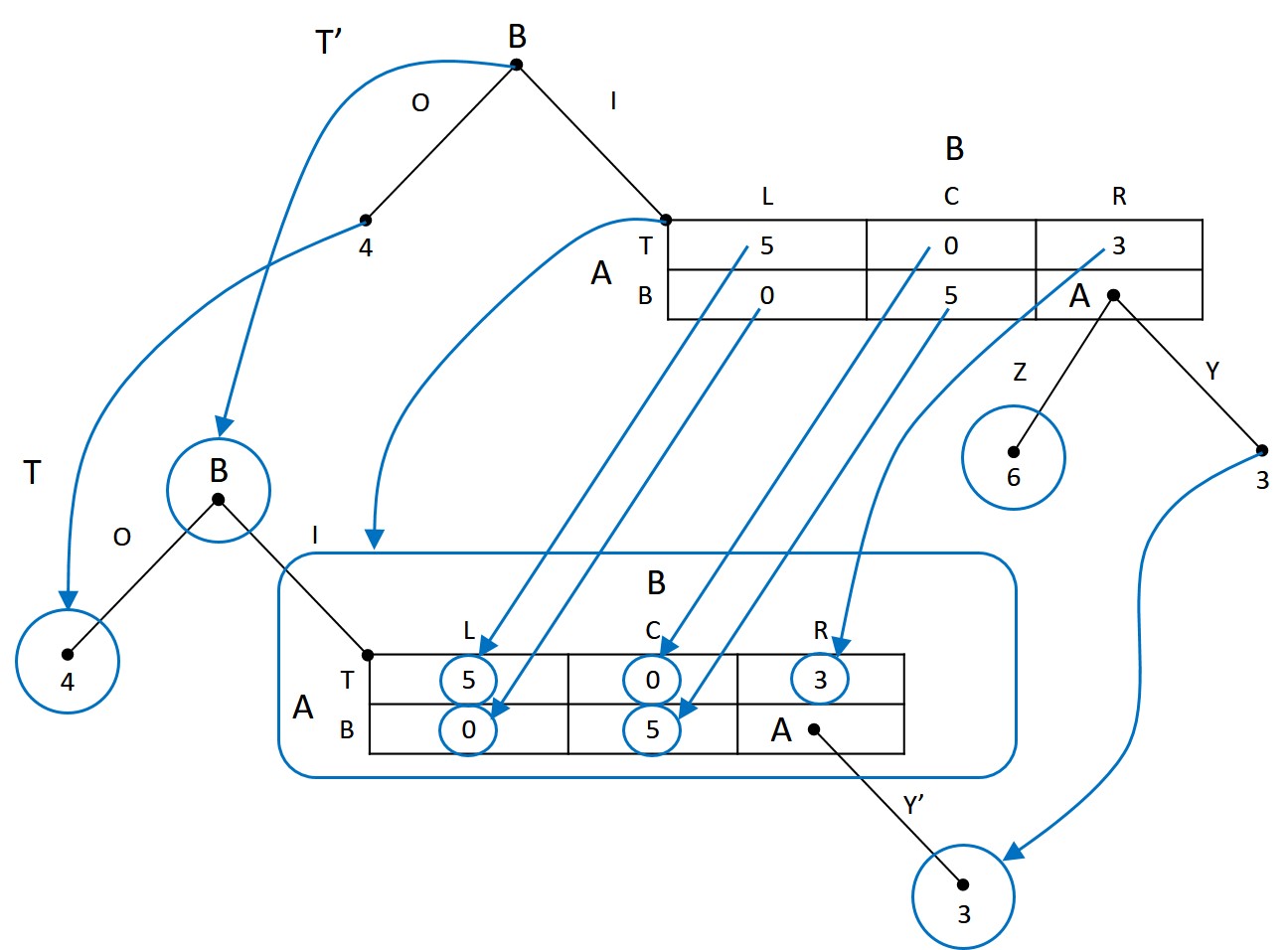}
\end{center}
\end{figure}

A reviewer kindly suggested the game in Figure~\ref{awareness_unawareness0} to illustrate and motivate the issues associated with awareness of unawareness. There are two players, player A and player B. Player B moves first and can decide whether or not end the game, in which case both players obtain 4, or to continue in which case both players choose simultaneously in a second stage. In that second stage, player A can decide between two actions, $T$ and $B$, whereas player B has three actions, $L$, $C$, and $R$. No matter what combination of actions is chosen by both players, the game ends except when player A chooses $B$ and player B chooses $R$. In that case, player A can choose again between $Z$ and $Y$ and the game ends. Initially, player B is unaware of player A's action Z. This is indicated by the blue arrows and ovals, player B's information sets. In contrast, player A is fully aware. For simplicity, we omit his information sets in Figure~\ref{awareness_unawareness0}. Since we assume that the game models common interest, the payoffs at terminal histories are the payoffs to either player.

In any extensive-form rationalizable outcome, player B won't play $IR$ since it is dominated initially by moving out, $O$. Thus, player B won't discover player A's action $Z$. In some sense, playing $R$ is attractive. It ``hedges'' the strategic uncertainty stemming from actions $T$ and $B$ of player A conditional of having moved in. Each player gets 3 for sure while (at least in player B's mind) while otherwise one has to face the strategic uncertainty of getting 5 or 0. More importantly, if player B somehow suspects that he could be unaware of something that is good for him, then he may want to experiment with playing $IR$. This becomes even more compelling if player B not just considers the payoff from playing the game once but also future repetition of the strategic situations. While strictly speaking these considerations are outside our model, we can at least sketch here how to extend our set up to awareness of unawareness.
\begin{figure}[h!]
\caption{\label{awareness_unawareness1}}
\begin{center}
\includegraphics[scale=0.45]{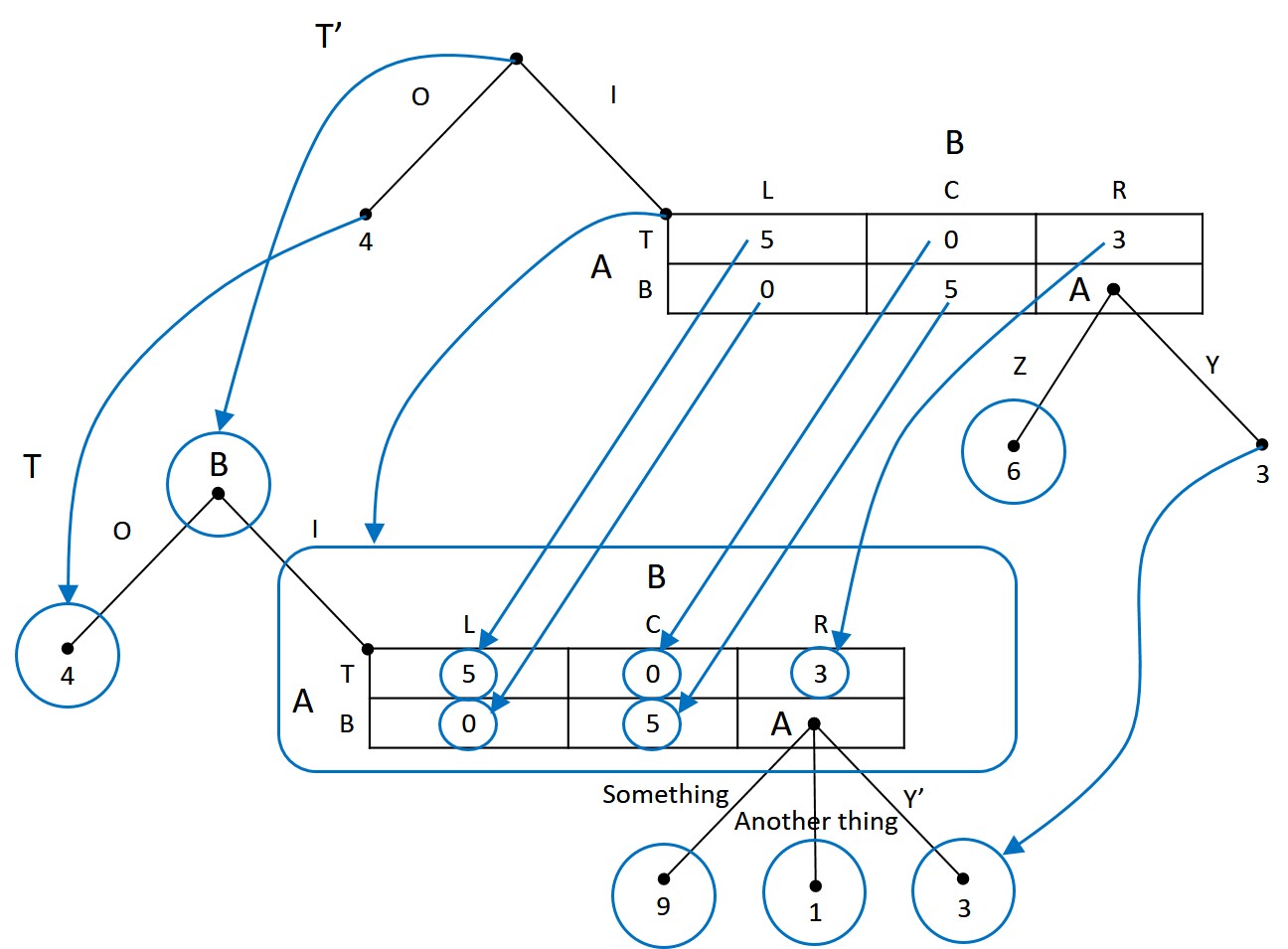}
\end{center}
\end{figure}

First, we require a the slightly more general set up of Heifetz, Meier, and Schipper (2013) allowing for imaginary moves representing vague ideas what could happen at some histories. For our example, the lower tree of Figure~\ref{awareness_unawareness1} represents the state of mind of player B who is aware that he is unaware of some moves of player A. He does not know the moves of player A but knows that there is ``something'' and perhaps ``another thing''. The payoffs at those imaginary terminal histories reflect that the that he must evaluate the terminal histories somehow.\footnote{In line with standard game theory, we do not present a theory of how these imaginary terminal histories are evaluated. The payoffs are primitives of the game.} Actions ``something'' and ``another thing'' in tree $T$ are both mapped to action $Z$ in tree $T'$. Obviously, the lower tree $T$ is not a subtree of $T'$ anymore. Yet, both trees could be subtrees of an even larger tree not shown here.
\begin{figure}[h!]
\caption{\label{awareness_unawareness2}}
\begin{center}
\includegraphics[scale=0.45]{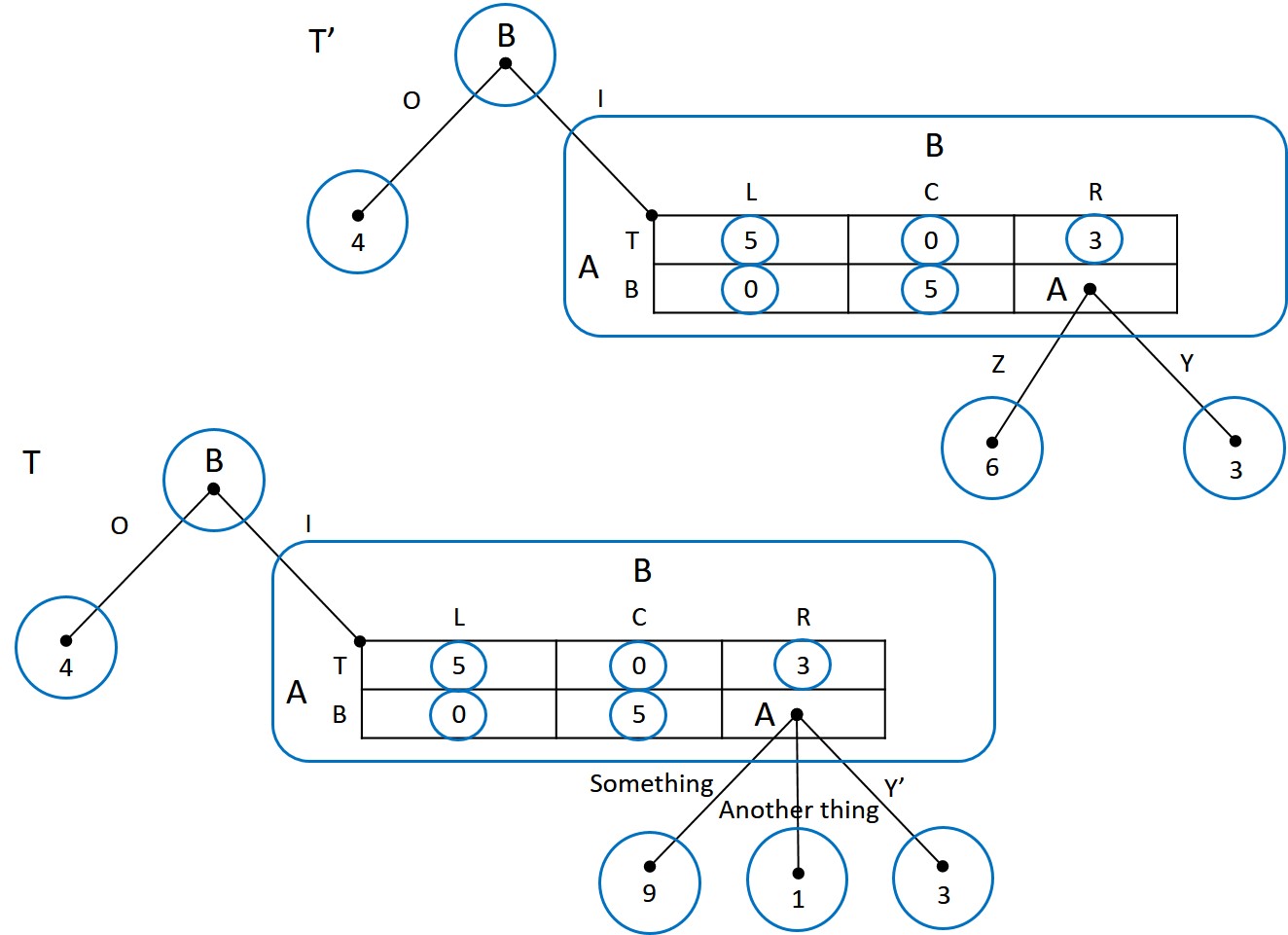}
\end{center}
\end{figure}

In the game of Figure~\ref{awareness_unawareness1}, moving $IR$ is now uniquely extensive-form rationalizable for player B because he anticipates that player A will take ``something'' upon which player B discovers history $Z$. Consequently, the ``next'' time the game is played, it is represented by the game in Figure~\ref{awareness_unawareness2}, the discovered version. This game differs from the prior game just in the updated information sets that follow the same principles as outline in the main section of this paper. In this discovered version, $IR$ remains uniquely rationalizable for player B. Consequently, it is part of the self-confirming equilibrium.

At this point, we leave a full development of awareness of unawareness and experimentation with the goal of discovering new actions to future research. Such a setting will also allow naturally for intertermporal maximization of payoffs across stage-games and lead to a ``repeated games'' notion of self-confirming equilibrium.

\subsection{Related Literature}

Obviously this paper builds upon Heifetz et al. (2013). In particular, we slightly extend the notion of generalized extensive-form games with unawareness and use their extension of extensive-form rationalizability. However, updating information upon discovery, the notions of discovered version of the game, discovery process, self-confirming games, and self-confirming equilibrium are new and are not even mentioned in Heifetz et al. (2013). Moreover, in Heifetz et al. (2013) we entirely rejected the notion of equilibrium to games with unawareness while in this paper I show how to extend the notion of self-confirming equilibrium so as to make sense under unawareness with repeated discoveries.

There is a large and growing literature on self-confirming equilibrium in games starting with Battigalli (1987), Fudenberg and Levine (1993a), and Kalai and Lehrer (1993b), who use the terminology ``conjectural'', ``self-confirming'', and ``subjective'' equilibrium, respectively.\footnote{See Oechssler and Schipper (2003) for a first experiment on self-confirming equilibrium in games.} In this literature, a mixed strategy of a player is either interpreted as a distribution of pure strategies in the population (e.g. Fudenberg and Levine, 1993a, Battigalli et al. 2015) or as a mixture played by a fixed set of players (e.g., Fudenberg and Kreps, 1995, Dekel, Fudenberg, and Levine, 1999, 2002). We prescribe to the latter interpretation because when considering a population of players for each player position, we would also have to specify a distribution of awareness for those players in the population, which would complicate the setting considerably. The focus on latter interpretation also justifies the use of what has been called unitary belief assumption (Fudenberg and Levine, 1993a). Instead of defining self-confirming equilibrium such that for each strategy in the support of the mixed equilibrium strategy there is a belief about opponents' strategies with which it is rational, we define it as there exist a belief over opponents' strategies for which every strategy in the support of the mixed equilibrium strategy is rational.

Various notions of rationalizable self-confirming equilibrium have been introduced in the literature. Battigalli and Guaitoli (1997) briefly discuss conjectural equilibrium in extensive-form rationalizable strategies with point-beliefs and pure strategies (and without unawareness) in a particular example of a macroeconomic game. In a recent paper most closely related to ours, Battigalli and Bordoli (2020) present a general exposition of self-confirming equilibrium in extensive-form rationalizable strategies in games without unawareness and provide a learning foundation in the spirit of Kalai and Lehrer (1993a). Rubinstein and Wolinsky (1994) consider rationalizable conjectural equilibrium in normal-form games (without unawareness) in which players receive signals about other player's actions and it is common knowledge that every player maximizes expected payoff given her signal. Gilli (1999) presents a strengthening of rationalizable conjectural equilibrium by Rubinstein and Wolinsky (1994). His refinement is equivalent to ours in games without unawareness when restricting to pure strategies. This follows from his Theorem 5. Esponda (2013) extends rationalizable conjectural equilibrium of Rubinstein and Wolinsky (1994) to static games with incomplete information (without unawareness). He also provides a characterization in terms of a procedure of iterative elimination of strategies as well as an epistemic characterization. Dekel, Fudenberg, and Levine (1999, 2002) introduce a notion of rationalizable self-confirming equilibrium for extensive-form games. Their notion differs from ours in several respects. First, they restrict players' beliefs to correspond to independent randomizations whereas we and Rubinstein and Wolinsky (1994) allow for correlation in the spirit of correlated rationalizability. Second, they assume that the path of play is public information to all players whereas we allow for more general information/signals. Third, our notion of rationalizable discovery process motivates restricting self-confirming equilibrium to extensive-form rationalizable strategies while in Dekel, Fudenberg, and Levine (1999, 2002) a player's strategy need to be optimal only at all of her information sets that are not precluded by the strategy itself. Finally, Dekel, Fudenberg, and Levine (1999, 2002) do not consider unawareness. Fudenberg and Kamada (2015, 2018) generalize the solution concept by Dekel, Fudenberg, and Levine (1999, 2002) to allow for more general information partitions over terminal nodes.

Self-confirming equilibrium has also been extended to contexts with more radical uncertainty such as ambiguity. Battigalli et al. (2015) study self-confirming equilibrium in the smooth ambiguity model. They show that in simultaneous games the set of self-confirming equilibria is expanding in ambiguity aversion. Moreover, an intuitive status-quo biases emerges: Ambiguity about played actions vanishes while ambiguity about unused actions remains present making latter even less attractive. Battigalli et al. (2018) show that these results do not extend to sequential games. Unawareness is distinct from ambiguity. Ambiguity is about lacking the ability to judge uncertainties probabilistically. Unawareness is about lacking even conception of some uncertainties.

Our approach is related to the literature on strategic interaction with misspecified models. Esponda and Pouzo (2016) consider interaction in normal-form games with players that form subjective distributions over their outcomes conditional on their own signal and action. Their models may be misspecified in the sense that their set of priors may not include the true distribution. In equilibrium, players play optimally given their belief and every subjective distribution in the support of their belief minimizes a distance to the true distribution. Our approach and their's have in common that we both in some sense extend the notion of self-confirming equilibrium to misspecified models. We focus explicitly on misspecifications due to unawareness while in their setting unawareness does not play any explicit role. Their approach is restricted to normal-form games while we explicitly consider general extensive-form games in which players may not be able to commit ex ante to contingent plans of actions due to unawareness. Moreover, we explicitly make use of the extensive-form structure to refine self-confirming equilibrium by extensive-form rationalizability (and hence forward induction) while they simply lack the necessary game theoretic structure to do this in their ``reduced-form'' statistical approach. While we admire the elegance of their statistical approach, we see value in exploring the consequences of unawareness as a particular source of misspecification in an explicit game theoretic approach.

There are a few papers that are closely related to ours as they all combine ideas of some sort of self-confirming equilibrium with a context that allows for lack of awareness. Greenberg, Gupta, and Luo (2009) study solution concepts related to both rationalizable self-confirming equilibrium and extensive-form rationalizability in strategic situations in which each player can live in their ``own game'' that could be widely unrelated. In contrast to their approach, we do not allow players to entertain delusions. They are just allowed to ``miss'' some features of the game due to unawareness. Greenberg, Gupta, and Luo's (2009) notion of path-mutual accepted course of actions corresponds roughly to our notion of self-confirming equilibrium in extensive-form rationalizable conjectures but in some sense players are interpreted to play the \emph{same own games} over and over again. There is no discovery of structural features of the underlying strategic situation, which we believe is a key component of our modeling approach. Sasaki (2017) observes that Nash equilibrium may fail to model a stable convention in games with unawareness. He proposes essentially a notion of self-confirming equilibrium and shows that it may fail to exist in some games with unawareness. He just considers normal-form games with unawareness though. \v{C}opi\v{c} and Galeotti (2006) study normal-form games with unawareness in which the awareness is determined endogenously in an equilibrium of conceptions and behavior, which is very similar to our notion of self-confirming equilibrium in a self-confirming game. Their analysis is confined to normal-form games while we consider more generally extensive-form games. They do not have the analogue to our notion of discovery process nor do they consider an analogue to self-confirming equilibrium in extensive-form rationalizable strategies. In an unpublished thesis, Camboni (2015) studies self-confirming equilibrium in single-person decision problems with unawareness. He requires as we do that awareness is self-confirming. Recently, Tada (2021) and Kobayashi, Sasaki, and Tada (2021) present follow-up work to our paper. Tada (2021) studies discovery processes in normal-form games with unawareness leading to CURB sets. Kobayashi, Sasaki, and Tada (2021) focus on rationalizable self-confirming equilibrium in normal-form games with unawareness. Note that normal-form games with unawareness are a special case of our framework as we allow for simultaneous moves. 

Somewhat further away from our approach is inductive game theory (e.g., Kaneko and Kline 2008). We share the desire to explain the origins of players' views of strategic situations. Moreover, both approaches feature a feedback between behavior and the players' views and the possibility of multiplicity of views. Yet, there are important conceptual differences. The authors assume that even initially some notion of recurrent situation and regular behavior exists while we would reject it in light of our introductory example. Moreover, they postulate that players experiment exogenously from time to time while we prescribe rationalizable behavior along the discovery process. Players can also forget infrequent features of the game in inductive game theory while our players' awareness can only (weakly) increase. From private communication I also know that Mamoru Kaneko rejects the use of extensive-form rationalizability. Yet, despite the differences of our formal approaches and conceptual differences with respect to solution concepts, we both share the aim of extending the tools of game theory so as to model more carefully the players' perceptions of strategic situations.

Our notion of discovered game can be understood as (an extensive-form) game theoretic analogue to awareness bisimulation in van Ditmarsch et al. (2018). Awareness bisimulation is used to compare awareness structures of Fagin and Halpern (1988). Roughly, two awareness structures are awareness bisimilar if they model the same information modulo awareness. In our context, the discovered version of a game with unawareness models the same information modulo awareness.

We view our paper adding the equilibrium notion to the tool-box of game theory with unawareness. This notion should prove useful in applications.

\appendix

\section{Proofs}

\subsection*{Proof of Remark~\ref{existence}}

If $\Gamma$ is an extensive-form game with common constant awareness then there exists $T \in \mathbf{T}$ such that for all $n \in \bar{T}$, $h_i(n) \subseteq T$ for all $i \in I$. This tree together with the information sets of all players on this tree constitutes an extensive-form game without unawareness. Since it is finite, it follows now from Nash's existence theorem and Kuhn's theorem on the equivalence of behavior strategies and mixed strategies in extensive-form games with perfect recall (and without unawareness) that the game possesses a Nash equilibrium in behavior strategies. Let $\pi$ denote a Nash equilibrium in behavior strategies of this game. Construct a profile of belief systems $\mu$ such that for every $i \in I$, $\mu_i(h_i)(\{\pi_{-i}\}) = 1$ for $h_i \in \tilde{H}_i(\pi)$ such that there is no $h'_i \in \tilde{H}_i(\pi)$ with $h_i' \rightsquigarrow h_i$. That is, $h_i$ is player $i$'s initial information set along the path induced by $\beta$. Consistent with conditioning of belief systems, $\mu_i(h''_i)(\{\pi_{-i}\}) = 1$ for all $h''_i \in \tilde{H}_i(\pi)$. For other information sets that are not on Nash equilibrium paths, we let beliefs to be arbitrary except that we require them to satisfy properties of a belief system. We claim that $\pi$ is a self-confirming equilibrium in behavior strategies: Condition (0) is vacuously satisfied in finite extensive-form games without unawareness. Condition (i) is implied by $\pi$ being a Nash equilibrium. For any player $i \in I$, given belief system $\mu_i$, Nash equilibrium behavior strategy $\beta_i$ maximizes player $i$'s expected payoff at every of her (non-terminal) information sets along the equilibrium paths. Condition (ii) is satisfied by construction of $\mu$. Now extend the equilibrium to a self-confirming equilibrium of $\Gamma$. Such an extension is possible because conditions of self-confirming equilibrium pertain to beliefs and information sets along the equilibrium path in $T$ only, for all $i \in I$. To see that the converse does not hold, consider for instance the game in Figure~\ref{example2b} as a counterexample. \hfill $\Box$

\subsection*{Proof of Proposition~\ref{well-defined}\label{proof_well-defined}}

We show that for any extensive-form game with unawareness $\Gamma \in \mathbf{\Gamma}$ and any strategy profile $s_{\Gamma} \in S_{\Gamma}$, the discovered version $\Gamma'$ satisfies all properties 1--3, U0, U1, U4, U5, and I2-I7. Properties 1--3 as well as I7 have been listed in the main-text. For easy reference, we list the remaining properties (see Heifetz, Meier, and Schipper, 2013):
\begin{itemize}
\item[U0] Confined awareness: If $n \in T$, then there exists $T' \in \mathbf{T}$ with $h_i(n) \subseteq T'$ and $T' \preceq T$.

\item[U1] Generalized reflexivity: If $T' \preceq T$, $n \in T$, $h_i(n) \subseteq T'$ and $T'$ contains a copy $n_{T'}$ of $n$, then $n_{T'} \in h_i(n)$.

\item[I2] Introspection: If $n' \in h_i(n)$, then $h_i(n') = h_i(n)$.

\item[I3] No divining of currently unimaginable paths, no expectation to forget currently conceivable paths: If $n' \in h_i(n) \subseteq T'$ (where $T' \in \mathbf{T}$ is a tree) and there is a path $n', \dots , n'' \in T'$ such that $i \in P(n') \cap P(n'')$, then $h_i(n'') \subseteq T'$.

\item[I4] No imaginary actions: If $n' \in h_i(n)$, then $A_{n'}^i \subseteq A_n^i$.

\item[I5] Distinct action names in disjoint information sets: For a subtree $T \in \mathbf{T}$, if there a decision nodes $n, n' \in T \cap \mathbf{D}$ with $A_n^i = A_{n'}^i$, then $h_i(n') = h_i(n)$.

\item[I6] Perfect recall: Suppose that player $i$ is active in two distinct nodes $n_{1}$ and $n_{k}$, and there is a path $n_{1}, n_{2}, ..., n_{k}$ such that at $n_{1}$ player $i$ takes the action $a_{i}$. If $n^{\prime} \in h _{i}\left( n_{k}\right)$, then there exists a node $n_{1}^{\prime}\neq n^{\prime }$ and a path $n_{1}^{\prime}, n_{2}^{\prime }, ..., n_{\ell}^{\prime } = n^{\prime }$ such that $h_{i}\left( n_{1}^{\prime}\right) = h_{i} \left( n_{1}\right) $ and at $n_{1}^{\prime }$ player $i$ takes the action $a_{i}$.

\item[U4] Subtrees preserve ignorance: If $T \preceq T' \preceq T''$, $n \in T''$, $h_{i}(n) \subseteq T$ and $T'$ contains the copy $n_{T'}$ of $n$, then $h_{i}(n_{T'}) = h_{i}( n)$.

\item[U5] Subtrees preserve knowledge: If $T \preceq T' \preceq T''$, $n \in T''$, $h_{i}(n) \subseteq T'$ and $T$ contains the copy $n_{T}$ of $n$, then $h_{i}(n_{T})$ consists of the copies that exist in $T$ of the nodes of $h_{i}(n)$.
\end{itemize}

For any player $i \in I$, we verify these properties one-by-one.

Properties 1--3: This follows directly from Definition~\ref{discovered_version} (i) and the fact that $\Gamma$ satisfies properties 1--3.

U0: Definition~\ref{discovered_version} (ii a.) says that for any $n \in T''$, the redefined information set is a subset of $T_{s_{\Gamma}}^i$ with $T_{s_{\Gamma}}^i \preceq T''$. Definition~\ref{discovered_version} (ii b.) says that for any $n \in T''$, the redefined information set is a subset of $T''$. All other information sets (Definition~\ref{discovered_version} (ii c.)) remain unchanged. Thus, $\Gamma'$ satisfies U0 because $\Gamma$ does.

For the proof of the remaining properties, we use the fact that $\Gamma'$ satisfies U0 without always explicitly mentioning it.

U1: Consider first information sets redefined by Definition~\ref{discovered_version} (ii a.). By assumption $\bar{T} \succeq T'' \succeq T^i_{s_\Gamma} \succeq T'$, $n \in \bar{T}$, $h_i(n) \subseteq T'$. If $n_{T^i_{s_\Gamma}}$ is the copy of $n$ in $T^i_{s_\Gamma}$, then since $\Gamma$ satisfied U4 we have $h_i(n_{T^i_{s_\Gamma}}) = h_i(n)$. Since $n_{T^i_{s_\Gamma}} = (n_{T''})_{T^i_{s_\Gamma}}$, we have $h_i((n_{T''})_{T^i_{s_\Gamma}}) = h_i(n)$. Hence by Definition~\ref{discovered_version} (ii a.) $(n_{T''})_{T^i_{s_\Gamma}} \in h'_i(n_{T''})$.

Similarly, consider now information sets redefined by Definition~\ref{discovered_version} (ii b.). We claim that for all $T'' \in \mathbf{T}$ with $T' \preceq T'' \preceq T^i_{s_\Gamma}$, $n' \in T''$, and $i \in P'(n')$, we have $n' \in h'_i(n') \subseteq T''$ (which implies U1 for those information sets). By Definition~\ref{discovered_version} (ii b.), $n_{T''} \in h_i'(n_{T''})$ if $h_i(n_{T''}) = h_i(n)$. (Recall that $n \in \bar{T}$ and $h_i(n) \subseteq T'$.) Let $n' = n_{T''}$. Then $h_i(n_{T''}) = h_i(n)$ follows from $\Gamma$ satisfying U4.

Any other information set of $\Gamma$ on $T' \not\preceq T_{s_{\Gamma}}^i$ remains unchanged in $\Gamma'$ by Definition~\ref{discovered_version} (ii c.). Since $\Gamma$ satisfies U1, so do these information set in $\Gamma'$.

I2: Consider first information sets redefined by Definition~\ref{discovered_version} (ii a.). First we show if for some $n' \in \bar{T}$, $n'_{T^i_{s_\Gamma}} \in h'_i(n_{T''})$, then $h'_i(n'_{T^i_{s_\Gamma}}) \subseteq h'_i(n_{T''})$. (Assuming that there exists a $n' \in \bar{T}$ whose copy in $T^i_{s_\Gamma}$ is w.l.o.g. because any node in a lower tree must be the copy of a node in a larger tree.) $\tilde{n} \in h'_i(n'_{T^i_{s_\Gamma}})$ if and only if $h_i(\tilde{n}) = h_i(n')$, $\tilde{n} \in T^i_{s_\Gamma}$, by Definition~\ref{discovered_version} (ii a.). Since $\Gamma$ satisfies U4, $h_i(n') = h_i(n'_{T^i_{s_\Gamma}})$. Since $n'_{T^i_{s_\Gamma}} \in h'_i(n_{T''})$, $h_i(n'_{T^i_{s_\Gamma}}) = h_i(n)$ by Definition~\ref{discovered_version} (ii a.). Thus, $h_i(n') = h_i(n)$. Hence $h_i(\tilde{n}) = h_i(n)$. Therefore $\tilde{n} \in h'_i(n_{T''})$.

Second, we show that if for some $n' \in \bar{T}$, $n'_{T^i_{s_\Gamma}} \in h'_i(n_{T''})$, then $h'_i(n'_{T^i_{s_\Gamma}}) \supseteq h'_i(n_{T''})$. $\tilde{n} \in h'_i(n_{T''})$ if and only if $h_i(\tilde{n}) = h_i(n)$, $\tilde{n} \in T^i_{s_\Gamma}$ by Definition~\ref{discovered_version} (ii a.). By the same argument as for the proof of the converse above, U4 of $\Gamma$ implies $h_i(\tilde{n}) = h_i(n')$. Thus, $\tilde{n} \in h'_i(n'_{T^i_{s_\Gamma}})$.

The proof for information sets redefined in Definition~\ref{discovered_version} (ii b.) is analogous. Finally, all information sets defined in Definition~\ref{discovered_version} (ii c.) remained unchanged from $\Gamma$. Since $\Gamma$ satisfies I2, so do these information sets.

I3: We need to show that if $n_1 \in h'_i(n) \subseteq T^*$ and there is a path $n_1, ..., n_k \in T^*$ such that $i \in P(n_1) \cap P(n_k)$, then $h'_i(n_k) \subseteq T^*$. Consider first the case in which $h'_i(n) \subseteq T'$ and $T' \preceq T_{s_\Gamma}^i \preceq T'' \preceq \bar{T}$. By Definition~\ref{discovered_version} (ii a.), $h'_i(n_{T''}) = \{n' \in T^i_{s_\Gamma} : h_i(n') = h_i(n) \}$. Suppose now to the contrary that $n_1 \in h'_i(n_{T''})$ and there is a path $n_1, ..., n_k \in T^i_{s_\Gamma}$ such that $i \in P(n_1) \cap P(n_k)$ and $h'_i(n_k) \subseteq T''' \not\preceq T^i_{s_\Gamma}$. Such information set $h'_i(n_k)$ violates U0. Hence $T''' \preceq T^i_{s_\Gamma}$. But then $h'_i(n_k)$ violates Definition~\ref{discovered_version} (ii. a).

Consider now the case in which $h'_i(n) \subseteq T'$ and $T' \preceq T'' \preceq T^i_{s_\Gamma}$. By Definition~\ref{discovered_version} (ii b.), $h'_i(n_{T''}) = \{n' \in T'' : h_i(n') = h_i(n) \}$. Information sets are now stationary in the sense that the node at which the information occurs is an element of the information set. Thus, information sets in such trees satisfy I3.

Finally, for all other cases, I3 follows from the fact that $\Gamma$ satisfies I3.

I4: This follows immediately from $\Gamma$ satisfying I4 and $\Gamma'$ satisfying U0.

I5: This follows immediately from Definition~\ref{discovered_version} and the fact that the original game $\Gamma$ satisfies I5. Note that equivalent information sets are treated equally in Definition~\ref{discovered_version}.

I6: Suppose there is a path $n_1, n_2, ... n_k$, $n_1 \neq n_k$, with $i \in P(n_1) \cap P(n_k)$ such that at $n_1$ player $i$ takes action $a_i$. Moreover, let $n' \in h'_i(n_k)$ with $n' \neq n_k$. Further, suppose now to the contrary that there doesn't exist $n'_1 \neq n'$ with a path $n'_1, n'_2, ..., n'_\ell = n'$ such that $h'_i(n'_1) = h'_i(n_1)$ and at $n'_1$ player $i$ takes action $a_i$.

Consider first the case in which $h'_i(n_k) \subseteq T'' \preceq T^i_{s_\Gamma}$. By Definition~\ref{discovered_version} (ii b.) we must have $n_1, n_2, ..., n_k \in T''$. Since $T'' \preceq \bar{T}$, there exists $m_k \in \bar{T}$ such that $(m_k)_{T''} = n_k$. By Definition~\ref{discovered_version} (ii b.), $h'_i(n_k) = h'_i((m_k)_{T''}) = \{\tilde{n} \in T'': h_i(\tilde{n}) = h_i(m_k)\}$. Moreover, there must be a path $m_1, ..., m_k \in \bar{T}$ with $(m_1)_{T''} = n_1$ and $(m_k)_{T''} = n_k$. Assume w.l.o.g. that $T'' \prec \bar{T}$. Otherwise, if $T'' = \bar{T}$, then $\bar{T} = T_{s_{\Gamma}} = T''$ and Definition~\ref{discovered_version} leaves information sets unchanged in these cases. From $h_i(\tilde{n}) = h_i(m_k)$ we conclude by U1 of $\Gamma$ that $\tilde{n} \in h_i(m_k)$ with $\tilde{n} \neq m_k$. By I6 of $\Gamma$, there exists $\tilde{n}_1 \neq \tilde{n}$ and a path $\tilde{n}_1, \tilde{n}_2, ..., \tilde{n}_\ell = \tilde{n}$ (all in $T''$) such that $h_i(\tilde{n}_1) = h_i(m_1)$ and at $\tilde{n}_1$ player $i$ takes action $a_i$. By Definition~\ref{discovered_version} (ii b.), $h'_i(\tilde{n}_1) = \{n' \in T'' : h_i(n') = h_i(m_1)\} = h'_i((m_1)_{T''}) = h'_i(n_1)$, a contradiction in this case.

The remaining cases, $h'_i(n_k) \subseteq T'' \not\preceq T^i_{s_\Gamma}$, follow directly from I6 of $\Gamma$.

I7: This follows from from Definition~\ref{discovered_version} and the fact that the original game $\Gamma$ satisfies I7. Note that information sets of the original game consisting of terminal nodes give rise to information sets of terminal nodes in the discovered version. To show that if $z' \in h'_i(z)$ then $u_i(z') = u_i(z)$, suppose to the contrary that $u_i(z') \neq u_i(z)$. Then by Definition~\ref{discovered_version} there exists $z' \in h'_i(z)$ such that $h_i(z') = h_i(z^*)$ with $(z^*)_{T_{h'_i(z)}} = z$. By U1, if $h_i(z^*) \subseteq \hat{T}$ then $(z^*)_{\hat{T}} \in h_i(z^*)$ and by Definition~\ref{discovered_version} (ii a., b.) $(z')_{\hat{T}} \in h_i(z')$. Since the original game $\Gamma$ satisfies I7, $u_i((z')_{\hat{T}}) = u_i((z^*)_{\hat{T}})$ and thus $u_i(z') = u_i(z^*)$ by definition of $\Gamma$, a contradiction.

U4: First, consider the case $T^i_{s_\Gamma} \preceq T'' \preceq \bar{T}$, $n \in \bar{T}$, $h'_i(n) \subseteq T^i_{s_\Gamma}$, and $n_{T''} \in T''$. We need to show that $h'_i(n_{T''}) = h'_i(n)$. If $h'_i(n) = h_i(n)$, then the claim follows from U4 of $\Gamma$. Otherwise, if $h'_i(n) \neq h_i(n)$, then $h_i(n) \subseteq T'$ with $T' \preceq T^i_{s_\Gamma}$ by Definition~\ref{discovered_version} (ii a.). By Definition~\ref{discovered_version} (ii a.), $h'_i(n_{T''}) = \{n' \in T^i_{s_\Gamma} : h_i(n') = h_i(n) \}$. Thus, we need to show that $h'_i(n) = \{n' \in T^i_{s_\Gamma} : h_i(n') = h_i(n) \}$. But this follows Definition~\ref{discovered_version} (ii a.), i.e., $h'_i(n_{\bar{T}}) = \{n' \in T^i_{s_\Gamma} : h_i(n') = h_i(n) \}$ since $n = n_{\bar{T}}$.

Next, consider the case $T' \preceq T'' \preceq T^i_{s_\Gamma}$, $n \in T^i_{s_\Gamma}$, $h'_i(n) \subseteq T'$ and $n_{T''} \in T''$. We need to show that $h'_i(n_{T''}) = h'_i(n)$. By Definition~\ref{discovered_version} (ii b.), $h'_i(n_{T''}) \subseteq T''$ and $h'_i(n) \subseteq T^i_{s_\Gamma}$. Hence, $T^i_{s_{\Gamma}} = T' = T''$ and $h'_i(n_{T''}) = h'_i(n)$.

Finally, for any other case the information sets remain unchanged and the result follows from U4 of $\Gamma$.

U5: First, consider the case $T^i_{s_\Gamma} \preceq T'' \preceq \bar{T}$, $n \in \bar{T}$, $h'_i(n) \subseteq T''$, $n_{T^i_{s_\Gamma}} \in T^i_{s_\Gamma}$. We need to show that $h'_i(n_{T^i_{s_\Gamma}})$ consists of copies of nodes in $h'_i(n)$ in tree $T^i_{s_\Gamma}$. By Definition~\ref{discovered_version} (ii a.), $h'_i(n_{T^i_{s_\Gamma}}) = \{n' \in T^i_{s_\Gamma} : h_i(n') = h_i(n) \}$ and $h'_i(n) = \{n' \in T^i_{s_\Gamma} : h_i(n') = h_i(n) \}$. Hence $T'' = T^i_{s_\Gamma}$ and the claim follows.

Next, consider the case $T' \preceq T'' \preceq T^i_{s_\Gamma}$, $n \in T^i_{s_\Gamma}$, $h'_i(n) \subseteq T''$, $n_{T'} \in T'$. We need to show that $h'_i(n_{T'})$ consists of copies of nodes in $h'_i(n)$ in tree $T'$. By Definition~\ref{discovered_version} (ii b.), $h'_i(n_{T'}) = \{n' \in T' : h_i(n') = h_i(n)\}$ and $h'_i(n) = \{n' \in T^i_{s_\Gamma} : h_i(n') = h_i(n)\}$. Hence, $T^i_{s_\Gamma} = T''$. Since $\Gamma$ satisfies U5, $h'_i(n_{T'})$ consists of copies of nodes in $h'_i(n)$ in the tree $T'$ as required.

Finally, for any other case the information sets remain unchanged and the result follows from U5 of $\Gamma$.

To see that $\Gamma'$ has more awareness than $\Gamma$, note that from Definition~\ref{discovered_version} it is clear that if at some node an information set in the discovered version differs from the original game, then it is because it is has been raised to a more expressive tree.

Uniqueness follows from $T^i_{s_{\Gamma}}$ being unique for each $s_{\Gamma}$ and player $i \in I$ and the fact that Definition~\ref{discovered_version} uniquely redefines information sets of $\Gamma'$.

That $\Gamma'$ preserves information of $\Gamma$ follows from Definition~\ref{discovered_version}, the fact that $\Gamma'$ has more awareness than $\Gamma$, and the fact that both games satisfy U0, U1, and I2.

This completes the proof of Proposition~\ref{well-defined}. \hfill $\Box$


\subsection*{Proof of Theorem~\ref{main_theorem}}

By Proposition~\ref{existence_rscv} every extensive-form game with unawareness possesses a rationalizable self-confirming version. What is left to show is that every self-confirming game possesses a self-confirming equilibrium in extensive-form rationalizable conjectures.

Preliminary observation: Fix a finite extensive-form game with unawareness. By Remark 7 in Heifetz, Meier, and Schipper (2013), for every player $i \in I$, $R_i^k \subseteq R_i^{k-1}$ for $k > 1$. Since the game is finite, there exists $\bar{k}_i$ such that for all $k > \bar{k}_i$, $R_i^k = R_i^{\bar{k}_i}$. Thus, $R_i^\infty = R_i^{\bar{k}_i}$. Let $\bar{k} := \max_{i \in I} \{\bar{k}_i + 1\}$. From Definition~\ref{EFR} follows that $s_i \in R_i^{\infty}$ is rational for some $\beta_i \in B_i^{\bar{k}}$ at every information set of player $i$ \emph{among all} strategies in $S_i$ (not just among strategies in $R_i^{\bar{k}}$). More precisely, for each $s_i \in R_i^{\infty}$, there exists a belief system $\beta_i \in B_i^{\bar{k}}$ such that $s_i$ is rational with $\beta_i$ at all information sets $h_i \in H_i$. Moreover, any strategy $s_i$ that for some $\beta_i \in B_i^{\bar{k}}$ is rational at all information sets $h_i \in H_i$ is contained in $R_i^{\infty}$.\footnote{In other words, the set of extensive-form rationalizable strategies is an extensive-form best-response set in the terminology of Battigalli and Friedenberg (2012).} By Proposition 1 in Heifetz, Meier, and Schipper (2013), $R_i^{\infty}$ is nonempty for every $i \in I$.

Fix a rationalizable discovery process that has ``full support'' on all extensive-form rationalizable strategies in each stage-game $\Gamma \in \bm{\Gamma}$. By Proposition~\ref{existence_rscv} it has an absorbing state, a rationalizable self-confirming game $\Gamma^*$. Since it is an absorbing state of the rationalizable discovery process with ``full support'' starting in the initial game $\Gamma^0$, we have that in $\Gamma^*$ for each player $i \in I$, there is a tree $T^i \in \mathbf{T}$ such that for all $h_i \in \tilde{H}_i(s)$, for all $s \in R^{\infty}_{\Gamma^*}$, we have $h_i \subseteq T^i$.

Denote by $U_i^{\bar{T}}(\sigma)$ player $i$'s expected payoff from the mixed strategy profile $\sigma$ in the upmost tree $\bar{T}$. (Fix nature's strategy (if any) in an arbitrary way.) Note that if $\sigma \in \Delta(R^{\infty}_{\Gamma^*})$, then $U_i^{T^i}(\sigma) = U_i^{\bar{T}}(\sigma)$ (since nothing is discovered with any $s \in R^{\infty}_{\Gamma^*}$ in $\Gamma^*$). Define $\sigma^* \in \Delta(R^{\infty}_{\Gamma^*})$ by for all $i \in I$, $U_i^{\bar{T^i}}(\sigma^*) \geq U_i^{\bar{T^i}}(\sigma_i, \sigma^*_{-i})$ for all $\sigma_i \in \Delta(R_{\Gamma^*, i}^{\infty})$. (This may be interpreted as a Nash equilibrium in mixed strategies restricted to extensive-form rationalizable strategies only (given nature's strategy if any).) From our preliminary observation follows that this holds for all $\sigma_i \in \Delta(S_{\Gamma^*, i})$ (and not just for all $\sigma_i \in \Delta(R_{\Gamma^*, i}^{\infty})$). Hence, $\sigma^*$ is a Nash equilibrium of $\Gamma^*$ (given nature's strategy if any). By Nash's existence theorem and perfect recall, such a Nash equilibrium exist. By Kuhn's Theorem for games with unawareness (Schipper, 2018), we can consider the equivalent behavior strategy profile $\pi^*$. This is a self-confirming equilibrium. (0) is implied by the fact that $\Gamma^*$ is a rationalizable self-confirming game and $\sigma^* \in \Delta(R^{\infty}_{\Gamma^*})$. (i) and (ii) are implied by $\sigma^*$ being Nash equilibrium of $\Gamma^*$. \hfill $\Box$


\end{document}